\documentclass[10pt,twoside,twocolumn,english,aps,prl,superscriptaddress,notitlepage]{revtex4-2}

\usepackage[utf8]{inputenc}
\usepackage[a4paper]{geometry}
\usepackage[english]{babel}
\usepackage{color,xcolor,soul}
\usepackage{verbatim,textcomp}
\usepackage{amstext,amsfonts,amsmath,amssymb}
\usepackage{graphicx}
\usepackage[calcwidth,explicit]{titlesec}
\usepackage[normalem]{ulem}
\usepackage[unicode=true,pdfusetitle, bookmarks=false, breaklinks=true,pdfborder={0 0 0},pdfborderstyle={},backref=false,colorlinks=true]{hyperref}
\hypersetup{colorlinks=true,citecolor=blue,linkcolor=black,urlcolor=blue}
\usepackage{siunitx}
\usepackage{soul}
\usepackage{float}
\usepackage{makecell}
\usepackage{enumerate}

\makeatletter\renewcommand\frontmatter@abstractwidth{\dimexpr\textwidth-2cm\relax}\makeatother

\titleformat{\section}{}{}{0pt}{}
\titleformat{\section}{\bfseries\sffamily\large\filcenter}{\thesection.}{0.2em}{#1}
\titlespacing{\section}{0pt}{0.2ex}{0.1ex}
\titleformat{\paragraph}[runin]{\normalfont\normalsize\bfseries}{}{0pt}{}
\titlespacing*{\paragraph}{0em}{0ex}{0.3em}[]

{}

\renewcommand\thesection{\Alph{section}}%\renewcommand\thesubsection{\Alph{section}\Arabic{subsection}}

\makeatletter\@addtoreset{paragraph}{section}\makeatother
\makeatletter\def\p@paragraph{}\makeatother

\AtBeginDocument{
\renewcommand{\ref}[1]{\autoref{#1}}

}

\begin{document}

\title{Quantifying the magnetic interactions governing chiral spin textures \\
using deep neural networks} % characteristics % in chiral multilayers
\author{Jian Feng Kong}
\email{kong\_jian\_feng@ihpc.a-star.edu.sg}
\thanks{These authors contributed equally}
\affiliation{Institute of High Performance Computing, Agency for Science, Technology \& Research (A*STAR), 138632 Singapore}
\author{Yuhua Ren}
\thanks{These authors contributed equally}
\affiliation{Department of Physics, National University of Singapore, 117551 Singapore}
\author{M.S. Nicholas Tey}
\affiliation{Institute of Materials Research \& Engineering, Agency for Science, Technology \& Research (A*STAR), 138634 Singapore}
\author{Pin Ho}
\affiliation{Institute of Materials Research \& Engineering, Agency for Science, Technology \& Research (A*STAR), 138634 Singapore}
\author{Khoong Hong Khoo}
\affiliation{Institute of High Performance Computing, Agency for Science, Technology \& Research (A*STAR), 138632 Singapore}
\author{Xiaoye Chen}
\email{chen\_xiaoye@imre.a-star.edu.sg}
\affiliation{Institute of Materials Research \& Engineering, Agency for Science, Technology \& Research (A*STAR), 138634 Singapore}
\author{Anjan Soumyanarayanan}
\email{anjan@nus.edu.sg}
\affiliation{Department of Physics, National University of Singapore, 117551 Singapore}
\affiliation{Institute of Materials Research \& Engineering, Agency for Science, Technology \& Research (A*STAR), 138634 Singapore}

\begin{abstract}
\noindent 
The interplay of magnetic interactions in chiral multilayer films gives rise to nanoscale topological spin textures, which form attractive elements for next-generation computing.
Quantifying these interactions requires several specialized, time-consuming, and resource-intensive experimental techniques.
Imaging of ambient domain configurations presents a promising avenue for high-throughput extraction of the parent magnetic interactions. 
Here we present a machine learning-based approach to determine the key interactions — symmetric exchange, chiral exchange, and anisotropy — governing chiral domain phenomenology in multilayers. 
Our convolutional neural network model, trained and validated on over 10,000 domain images, achieved $R^2 > 0.85$ in predicting the parameters and independently learned physical interdependencies between them.
When applied to microscopy data acquired across samples, our model-predicted parameter trends are consistent with independent experimental measurements.
These results establish ML-driven techniques as valuable, high-throughput complements to conventional determination of magnetic interactions, and serve to accelerate materials and device development for nanoscale electronics.
\end{abstract}
\maketitle

\noindent
%%%% Introduction & Setup
\section{Introduction\label{sec:intro}}

\paragraph{Motivation}
The advent of chirality in magnetic thin films has created a vast zoo of nanometre-scale spin textures, including spin spirals, stripes, skyrmions, and beyond \cite{Nagaosa.2013, Soumyanarayanan.2016, Back.2020, Goebel.2021}.
Their ambient stability, electrical malleability, and material compatibility with established fabrication techniques, has elicited growing technological interest \cite{Soumyanarayanan.2016, Fert.2017, Back.2020}, especially as elements for sustainable computing architectures \cite{Song.2020, Bourianoff.2018, Pinna.2018}. 
The formation of these spin textures, as well as their critical static and dynamic attributes are governed by the interplay of three key magnetic interactions \cite{Bogdanov.2001, Woo.2016, Soumyanarayanan.2017}. 
The exchange stiffness, $A$, characterizes uniform ferromagnetic order, the effective anisotropy $K_{\rm eff}$, describes its preferred orientation, and the interfacial Dzyaloshinskii-Moriya interaction (DMI, $D$), determines the extent of chirality \cite{Bogdanov.2001}. 
Quantifying these interactions is imperative both for elucidating the behaviour of spin textures, and for designing functional materials and devices to exploit their properties \cite{Fert.2017}. 

\paragraph{Experimental Challenge}
The experimental determination of these interactions typically involves several independent measurements.  
Firstly, $K_{\rm eff}$ can be obtained by combining magnetometry measurements across in-plane (IP) and out-of-plane (OP) sample orientations \cite{Johnson.1996}.
Next, as $A$ and $D$ relate to the symmetric and anti-symmetric components of spin-wave dispersion respectively, both can in principle be independently extracted via wavevector-resolved Brillouin light scattering (BLS) spectroscopy \cite{Kuepferling.2020, Boettcher.2021}. 
However, this approach is challenging for several reasons: (a) BLS is a specialized, resource-intensive technique; (b) its evaluation has implicit dependencies on two additional techniques; and (c) the measured $A$ can be ambiguous in films with sizable DMI or dipolar effects \cite{Boettcher.2023}.  
Alternatively, if $A$ is known, $D$ can be extracted from the measured asymmetry in domain wall (DW) propagation, driven by external electric or magnetic fields, albeit with several inter-dependencies \citep{Je.2013, Hrabec.2014, Kuepferling.2020}.
However, while $A$ can be estimated reliably for thicker films using the Bloch $T^{3/2}$ law for temperature-dependent magnetization \cite{Vaz.2008, Nembach.2015}, this method does not work well for the ultrathin (1\,nm) limit relevant to chiral multilayers \cite{Boettcher.2023}.
Overall, given the considerable challenges and complexities of direct experimental determination, a high throughput technique enabling one-stop-shop estimation of all three parameters would be of immense value.

\paragraph{Prevailing ML Approaches}
Since the magnetic parameters together determine the domain morphology, it should encapsulate material information.
However, while the forward problem of determining the equilibrium domain configuration using the input parameters is micromagnetically well-defined \cite{Donahue.1998,Vansteenkiste.2014}, the inverse problem of determining parent magnetic interactions from domain morphology is extremely challenging.
Unsurprisingly, prior attempts have been largely limited to quantifying a single feature, e.g. domain periodicity, and thereby estimating the ratio $D/A$ via linear regression \cite{MoreauLuchaire.2016, Woo.2016, Soumyanarayanan.2017, Kozlov.2020, Chen.2022}. 
The recent introduction of machine learning (ML) approaches in materials science provides considerable promise to address this inverse problem \cite{Carleo.2019, Schmidt.2019}. 
Several works have used ML techniques to recognize magnetic phases from domain configurations \cite{Iakovlev.2018, Singh.2019, Kwon.2019, SalcedoGallo.2020, Wang.2020}, and extend regression techniques in estimating parent magnetic parameter(s) \cite{Kawaguchi.2021, Mamada.2021, Talapatra.2022}. % Kwon.2019a, 
However, these approaches have largely utilized complete vectorial magnetization as inputs to train the ML model \cite{Kwon.2020}, which imposes considerable constraints on their broader experimental applicability. % \cite{Chen.2015}. 

\paragraph{Results Summary}
Here, we present an approach to simultaneously estimate the three key magnetic parameters -- $A$, $K_{\rm eff}$, and $D$ -- from the OP magnetization of chiral magnetic domain configurations.
To this end, we employ a supervised regression model using convolutional neural networks (CNN), and train it on micromagnetically simulated images that exclude information inaccessible to common microscopy tools. 
By learning physical features such as domain boundaries, the model performs well on simulated validation data ($R^2 > 0.85$). 
When tested on domain images obtained from a series of chiral multilayers, the model predicts parameter trends consistent with independent experimental estimates. This paves the way for high-throughput characterization of technologically relevant magnetic thin films.

\section{Methodology \& Training}

\begin{figure}[htb]
    \centering
    \includegraphics[width=1\linewidth]{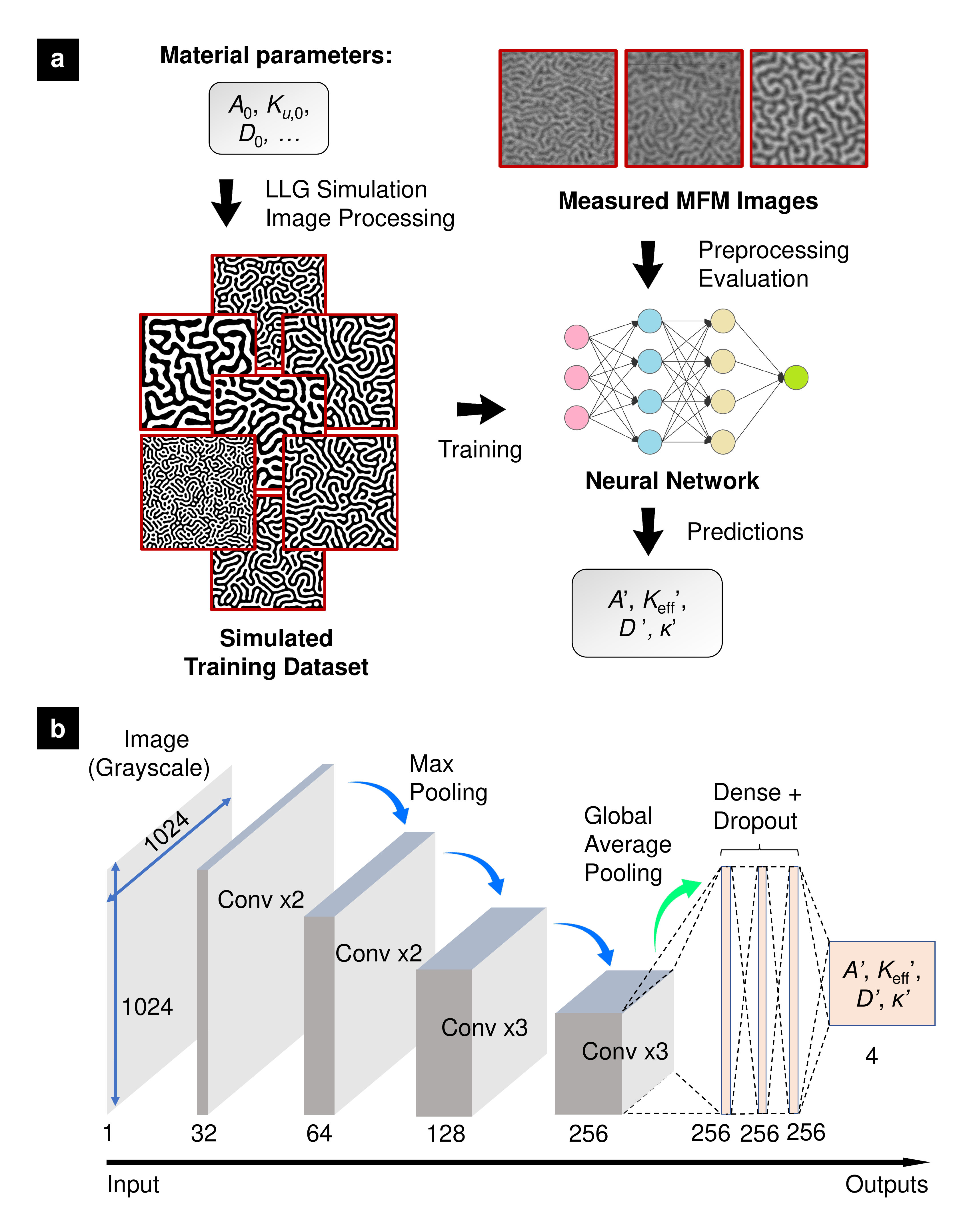}
    \caption{\textbf{Workflow and Model Architecture.}
    \textbf{(a)} Schematic of the training and prediction workflow. 
    Left: the model is trained using domain images generated by micromagnetic simulations (input parameters: $A_0, K_{\rm u, 0}, D_0$), post-processed to remove experimentally inaccessible information. 
    Right: The trained model is used to predict four magnetic parameters $(A', K
    _{\rm eff}, D', \kappa')$ on MFM-measured domain images. 
    \textbf{(b)} Schematic of the CNN model architecture, consisting of 10 convolutional layers, followed by 4 fully connected layers. The final layer is fully connected, with 4 neurons corresponding to the output parameters. See Methods for full details of the final model, including convolutional filters, max pooling layers, kernels, neurons etc.}
    \label{fig:architecture} 
\end{figure}

\paragraph{Model Choice \& Workflow}
The choice of ML model should optimize the learned relationship between input data features and desired outputs. % \cite{Alzubaidi.2021} 
For our purpose, a CNN model is suitable as it exploits many physical properties relevant to domain configurations \cite{LeCun.1989, LeCun.2015, Krizhevsky.2017}, such as translation invariance and local correlations arising from short-range exchange interactions.
In addition, CNN confers several practical benefits such as enabling variable-sized input images, and providing the potential for physical interpretability. % by inspection of intermediate feature maps. 
The schematic in \ref{fig:architecture}(a) outlines the ML workflow used in this work. 
First, a large set of ground-state domain images of OP magnetization was generated by performing micromagnetic simulations with varying input parameters. 
Next, these images were used to train our ML model, following post-processing to remove experimentally-inaccessible information. 
Finally, the model performance was evaluated by testing on an independent set of simulated images, and on experimental images acquired using a commercial magnetic force microscope (MFM). 

\paragraph{Training \& Data Generation}
An ML model requires a large volume of high-quality training data to reliably and effectively learn input-output correlations. 
In line with previous works, which used $\gtrsim 10^4$ domain configurations for training \cite{Kwon.2019, Kwon.2020, Kawaguchi.2021}, our CNN model was trained on a large dataset generated by micromagnetic simulations using mumax$^3$ \cite{Vansteenkiste.2014}.
Each of the input parameters, $A_0$, $K_{\rm u,0}$ ($K_{\rm u} = K_{\rm eff} + \mu_0 M_{\rm s}^2/2$, where $M_{\rm s}$ is the saturation magnetization), and $D_0$ was varied over a wide range for a multilayer comprising four repetitions of a chiral stack under the effective medium approximation \cite{Woo.2016}. 
The propensity for chiral domain formation can be characterized by the stability parameter, $\kappa$, defined as \cite{Bogdanov.2001, Soumyanarayanan.2017, Chen.2022}:
\begin{equation}
    \label{eqn:kappa}
    \kappa = \left. \pi D \right/ 4\sqrt{A K_{\rm eff}}.
\end{equation}
Within our simulations, multi-domain states form for $\kappa \gtrsim 0.1$, and proliferate with increasing $\kappa$ \cite{Chen.2022}. 
Meanwhile, the stripe width increases with $A$, $K_{\rm eff}$, and $D^{-1}$. 
Thus generated equilibrium magnetization configurations exhibiting multi-domain states were used for the training dataset. 

\begin{figure}
    \centering
    \includegraphics[width=0.9\linewidth]{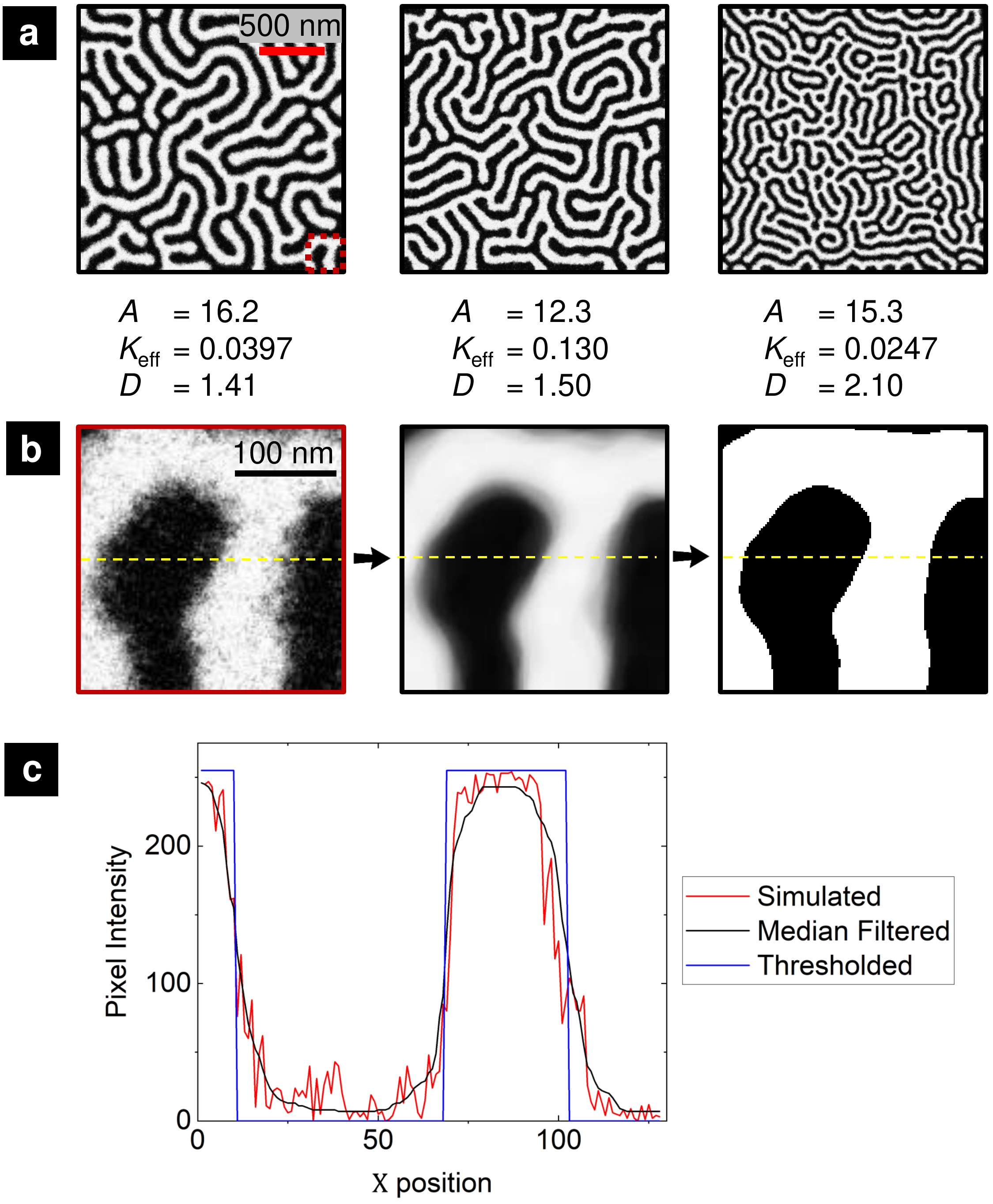}
    \caption{ \textbf{Simulated Images and Processing}. 
    \textbf{(a)} Sample images (grayscale: $m_z \in [-1,1]$) generated by micromagnetic simulations with varying input parameters ($A_0$ in \si{\pico\joule/\meter}, $D_0$ in \si{\mega\joule/\meter^3}, $K_{\rm{eff}}$ in \si{\milli\joule/\meter^2}).
    \textbf{(b)} Post-processing of simulated images for model training. 
    Left image: as-simulated from micromagnetics, including experimentally inaccessible domain wall (DW) and thermal field (fuzziness) information related to magnetic parameters. 
    Middle image: output of median filtering, which removes fuzziness within domains. 
    Right image: output of Otsu thresholding, wherein DW width information is removed.
    \textbf{(c)} Comparison of a representative line cut (dashed yellow line) across the three images in (b), showing the effects of the post-processing steps.}
    % , with the $z$-component of magnetization $m_z \in [-1,1]$ shown in grayscale
    % resulting in a sharp image with smooth domain boundaries,
    \label{fig:simImages}
\end{figure}

\paragraph{Data Filtering \& Augmentation}
To reliably predict outputs using commonly available magnetic microscopy techniques (e.g., MFM, Kerr microscopy), the ML model was trained only on the OP component, $m_z$, of normalized spatial magnetization, as exemplified in \ref{fig:simImages}(a).  
The training data were augmented by performing rotations on input domain images, consistent with the underlying symmetries of the problem \cite{Shorten.2019}.  
Our model is configured to predict three independent parameters -- $A'$, $K'_{\rm eff}$, and $D'$ -- and one dependent parameter, $\kappa'$.  
Note that $K'_\mathrm{eff}$ is used as the output parameter (instead of $K_{\rm u}$) to enable direct comparison with experiments.
Meanwhile, $\kappa'$ is included due to its physical significance \cite{Soumyanarayanan.2017, Chen.2022}, and to check whether the model is able to uncover the underlying constraints between the input parameters.
The CNN architecture employed in our work, shown in \ref{fig:architecture}(b), uses ReLU and leaky ReLU as non-linear activation functions for all intermediate layers, and includes dropout layers to reduce overfitting \citep{Hinton.2012}. 
The four max-pooling layers downsize the input by retaining the most relevant features in each patch, and are known to be effective in training CNN models \citep{Chollet.2017}. 

%%%%%%%%% Results & Impact

\section{Model Performance}

\begin{figure}[htb]
    \centering
    \includegraphics[width=0.9\linewidth]{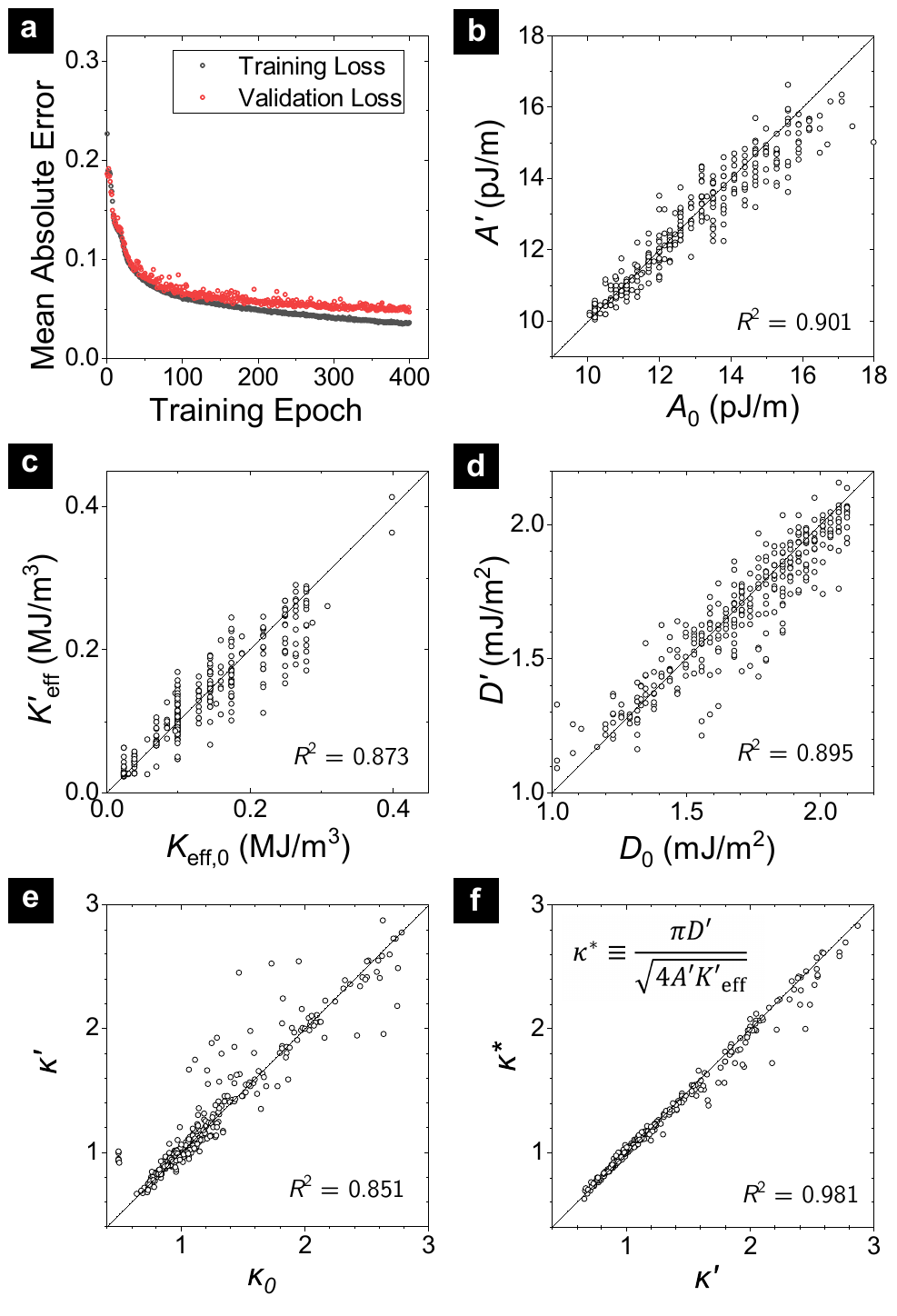}
    \caption{\label{fig:graphs} \textbf{Model Validation and Performance Testing. } 
    \textbf{(a)} Model loss during training and validation, quantified via mean absolute error (MAE) across training epochs. 
    \textbf{(b-e)} Parity plots for the trained model on the test dataset that compare input ($x$-axis) and predicted ($y$-axis) values for (b) $A$, (c) $K_{\rm eff}$, (d) $D$, and (e) $\kappa$. 
    The parity line ($y=x$ line) is shown for reference. %, and proximity to the line represents more accurate predictions. 
    \textbf{(f)} Parity plot for the predicted $\kappa'$ against $\kappa^* = \left. \pi D' \right/ 4\sqrt{A' K'_{\rm eff}}$ (\ref{eqn:kappa}) calculated from individually predicted parameters.} %, as a consistency check.
\end{figure}

\paragraph{Overfit Mitigation}
Allowing the model to train on as-simulated $m_z$ images produced unbelievably good initial validation results ($R^2 \geq 0.98$, SM \S S2), albeit with poor generalization to experimental data.   
This discrepancy arises from key differences in the short lengthscale ($< 10$\,nm) information encoded in the simulated images c.f. experimental images.
First, the DW width ($\sim 5$\,nm speckles), which relates to $\sqrt{A/K_{\rm u}}$ \cite{Vertesy.2003, Lemesh.2017}, is accessible within simulated images, but cannot be resolved in experiments (resolution $\sim 30$\,nm). 
Second, the thermal noise in mumax$^3$-simulated images ($\sim 5-10$\,nm), found to vary monotonically with $K_{\rm eff}$, may also leak experimentally-inaccessible information on the magnetic parameters  (SM \S S2). 
Meanwhile, the short lengthscale noise present in typical experimental images (\ref{fig:expt-setup}) is unrelated to magnetic properties. 
To enable accurate prediction for experimental images, we remove experimentally inaccessible data by judiciously pre-processing the simulated images, via median filtering and thresholding procedures (\ref{fig:simImages}(b-c)).  

\paragraph{Training Details}
The post-processed dataset of 12,000 simulated images was divided into 3 subsets for training (80\%), validation (10\%), and testing (10\%).  
Models with different hyperparameters (e.g., number of layers, filters, loss function, etc.) were individually trained, and subsequently, their performance was compared based on their predictions on the validation dataset (SM \S S1). 
The best-performing model thus identified was then evaluated using the testing dataset (\ref{fig:graphs}). 
In general, the models were found to learn more effectively with diminishing learning rates. %  (further details on CNN tuning in \cite{SSM})
\ref{fig:graphs}(a) shows the evolution of the training and validation losses for the chosen model during the training process. 
Both losses are found to decrease with the number of epochs trained, and the mean absolute error (MAE) asymptotes to a small, finite value after $\sim 400$ epochs. 
Crucially, the monotonic decrease of the validation loss indicates that the chosen model did not overfit the training data.

\paragraph{Validation Results}
\ref{fig:graphs}(b-e) summarizes the performance evaluation of chosen model on the testing dataset. 
The parity plots compare the predicted values for each of the output parameters, ($A',K'_{\rm eff},D',\kappa'$), with the respective input values, ($A_0,K_{\rm eff,0},D_0,\kappa_0$), provided for the simulations.
The consistently high $R^2$ for all determined parameters suggests that the ML model has successfully learned the intricate relationship between the domain configurations and input parameters.
As an additional test to examine the model's ability to learn implicit micromagnetic physics, we compare in \ref{fig:graphs}(f) the predicted kappa, $\kappa'$, with that resulting from \ref{eqn:kappa} via the predicted values of the other three parameters, $\kappa^* = \left. \pi D' \right/ 4\sqrt{A' K'_{\rm eff}}$.
The excellent correlation between $\kappa'$ and $\kappa^*$ ($R^2 > 0.98$) suggests that while the model does not receive \ref{eqn:kappa} as an explicit input, it is nevertheless able to accurately learn this implicit constraint during training, and implement it across predicted parameters.

\begin{figure}[htb]
    \centering
    \includegraphics[width=0.75\linewidth]{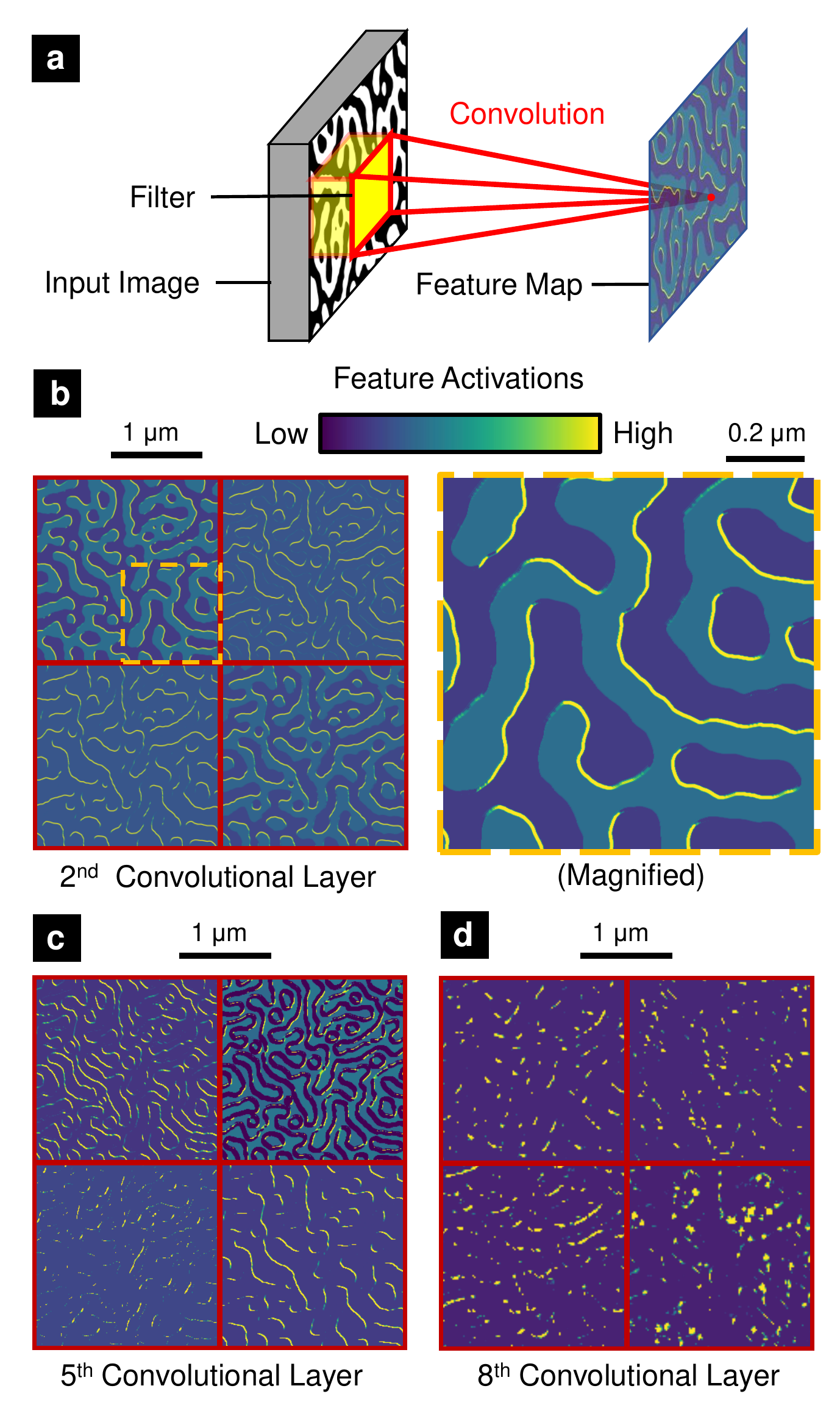}
    \caption{\label{fig:hidden} \textbf{Feature Maps. } 
    \textbf{(a)} Schematic of the convolution operation of CNN, and its correspondence to a feature map. 
    \textbf{(b-d)} Representative feature maps for the (b) 2nd (right inset: zoom-in), (c) 5th, and (d) 8th convolutional layers, with chosen filter numbers indicated. 
    Color represents activation strength.} % (individually normalized) 
\end{figure}

\paragraph{Feature Maps}
To visualize the functioning of the CNN, we examine the intermediate outputs of the convolutional filters -- known as feature maps (\ref{fig:hidden}(a)) -- as applied to the input. 
In practice, the CNN uses 1664 filters, and we display in \ref{fig:hidden}(b-d) a few representative feature maps that exemplify the pattern evolution across the neural network. 
While the outputs of the initial layers preserve the low-level spatial configurations from the input image, the deeper layers get progressively more abstract.
For example, the feature map for the 2nd layer (\ref{fig:hidden}(b)) clearly shows the DWs delineating regions of opposite magnetization. 
Moving to the 5th layer, the domains are still visible, but the orientation of magnetization is less clearly distinguished.
Finally, by the 8th layer, individual domains cannot be visually identified, and the observed speckles are abstract features used by the model to reach its final decision. 
Such behavior is consistent with the known progression of feature abstraction in deep neural networks \citep{Krizhevsky.2017}.

\section{Experimental Predictions}

\begin{figure}[htb]
	\centering
	\includegraphics[width=0.9\linewidth]{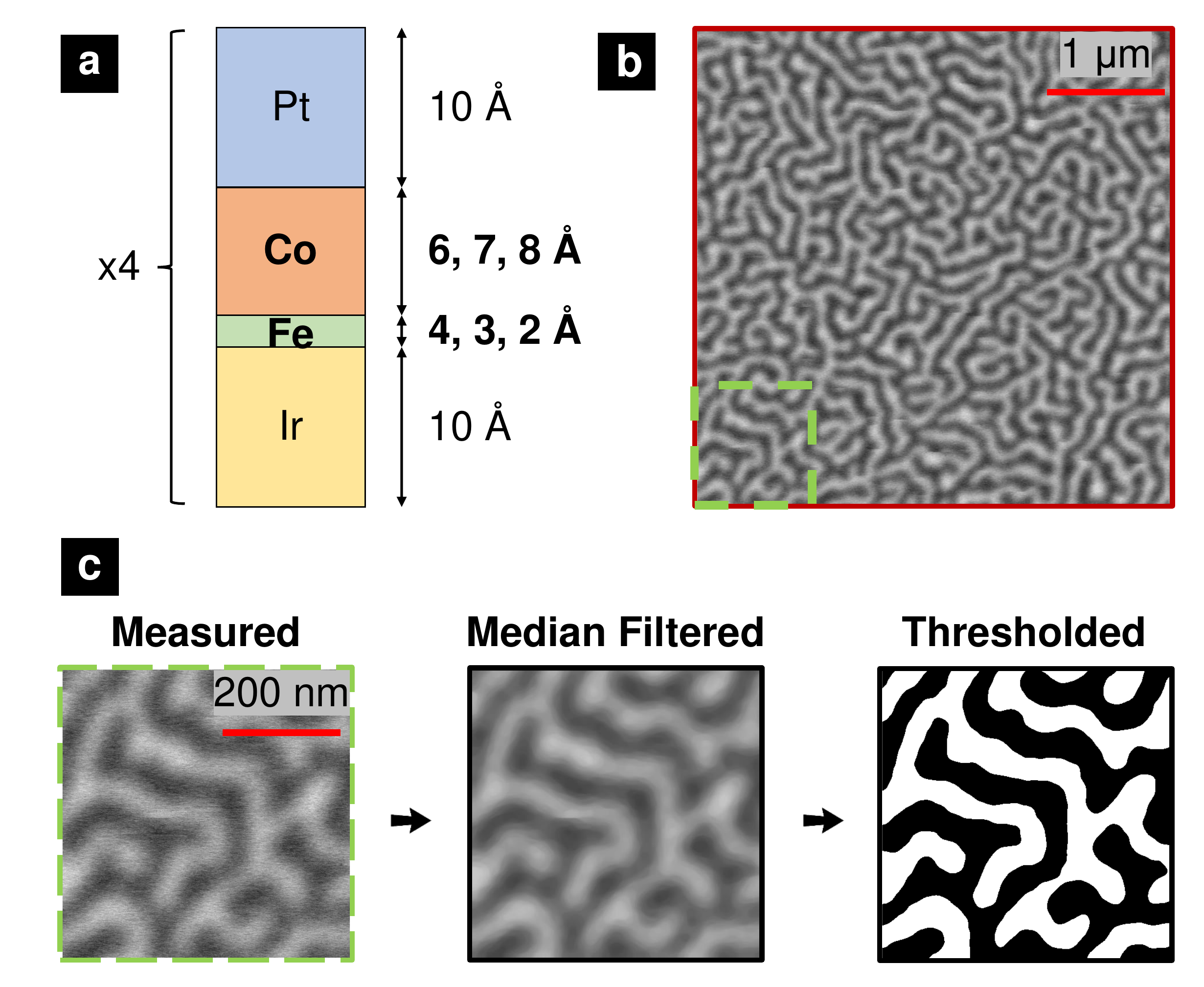}
	\caption{\label{fig:expt-setup} \textbf{Imaged Multilayer Samples.} 
    \textbf{(a)} Stack structure of Ir/Fe($x$)/Co($y$)/Pt samples. Varying Fe($x$)/Co($y$) thickness tunes the magnetic parameters. 
    \textbf{(b)} MFM-measured zero field domain image of Ir/Fe(2)/Co(8)/Pt sample. 
    \textbf{(c)} Processing of MFM-measured domain images for ML model testing. Left: measured MFM image (from (b): highlighted rectangle); centre: after median filtering; and right: after Otsu thresholding (see Methods).}
\end{figure}

\paragraph{Expt Prediction Setup}
Having comprehensively tested the trained model with simulated data, we proceed to evaluate its performance on experimentally acquired microscopic images from chiral multilayer films.
We use an established multilayer stack -- Ir(1)/Fe($a$)/Co($b$)/Pt (\ref{fig:expt-setup}(a)) -- wherein key magnetic parameters can be tuned by varying the Fe($a$)/Co($b$) thicknesses, while keeping their total thickness constant (1\,nm) \cite{Soumyanarayanan.2017, Chen.2022}.
The three samples studied in this work comprise four repetitions of stacks with varying Fe($a$)/Co($b$) compositions.  
The phenomenology and energetics of nanoscale equilibrium domain configurations ($\sim 100-300$\,nm, \ref{fig:expt-setup}(b)) in such DMI-dominated stacks ($D > 1$\,mJ/m$^2$) is qualitatively distinct from micron-scale domain configurations examined in prior ML works \cite{Kwon.2020, Kawaguchi.2021, Mamada.2021}. 
To image their domain morphology, we employ MFM, a commonly accessible, ambient, high-spatial-resolution ($\sim 30$\,nm) imaging technique, wherein measured contrast is proportional to the domain stray field gradient.  %  commonly accessible in materials science facilities  \cite{Kazakova.2019}
Close examination of a representative MFM image (\ref{fig:expt-setup}(c): left) reveals speckle noise and markedly reduced contrast c.f. simulated images. 
Therefore, MFM images are pre-processed following similar median filtering and thresholding procedures as used for the simulated images (\ref{fig:simImages}(b)). 
Post-processed MFM images (\ref{fig:expt-setup}(c): right) exhibit domain configurations similar to simulated images, which is further confirmed by corresponding MFM simulations (SI \S S3) 

\begin{figure}
    \centering
    \includegraphics[width=1\linewidth]{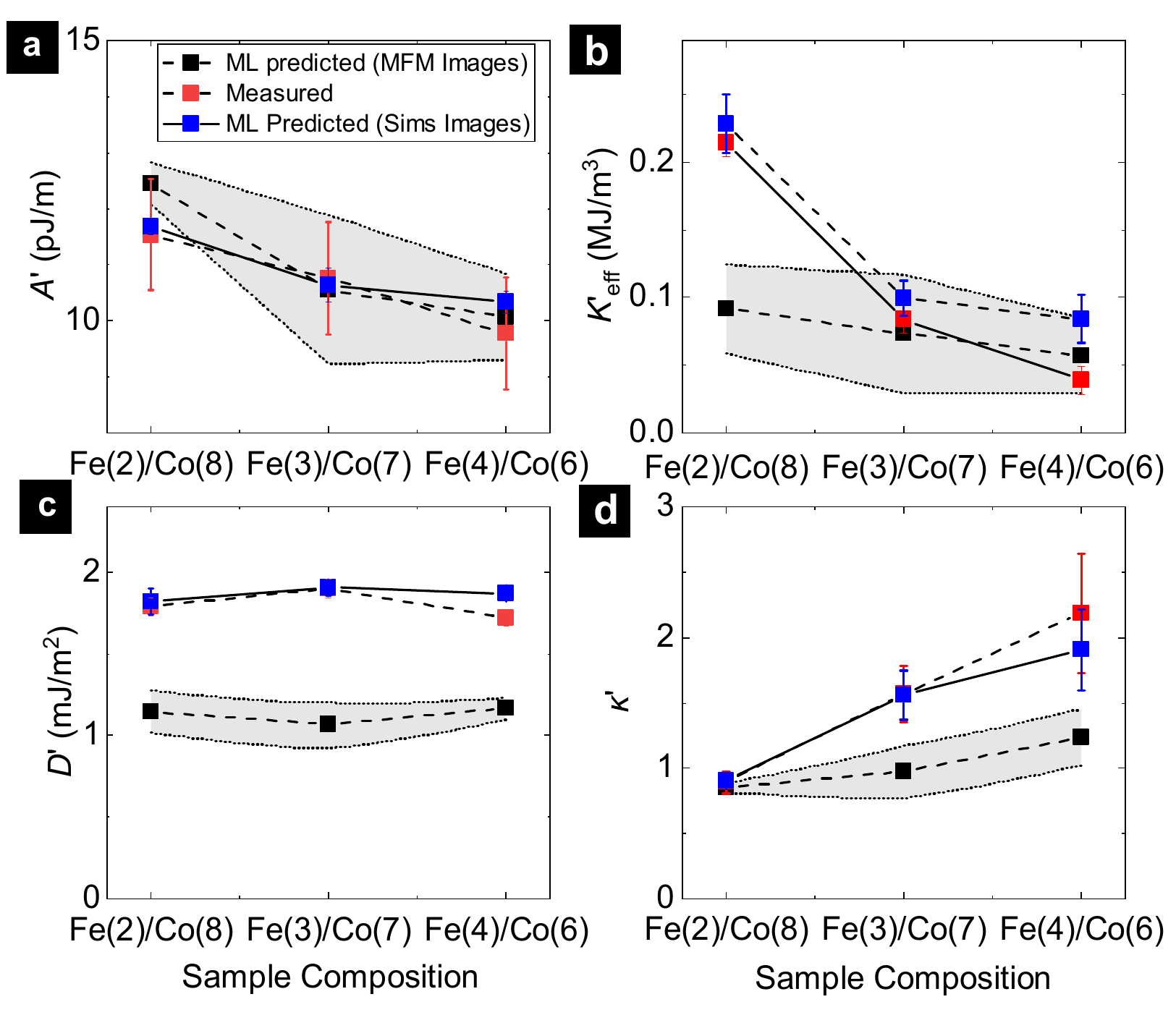}
    \caption{\label{fig:trends}\textbf{Measured \& Predicted Parameter Trends. } 
    Comparison of magnetic parameters -- (a) $A$ (b) $K_{\rm eff}$, (c) $D$, and (d) $\kappa$ -- for three Ir/Fe($x$)/Co($y$)/Pt samples. Plots show parameters obtained from experimental measurements (red), ML model prediction on measured MFM images (black), and ML model prediction on simulated images using measured parameters as inputs (blue).}
\end{figure}

\paragraph{Expt Prediction Results}
\ref{fig:trends} compares the parameters predicted by the ML model using experimentally acquired MFM images (black) on the samples with independently measured values (red). 
The latter set of values is obtained using a combination of magnetometry, BLS spectroscopy, and microscopy experiments, and are consistent with published results on these Fe($a$)/Co($b$) sample compositions \cite{Soumyanarayanan.2017, Chen.2022, Chen.2022a, Boettcher.2023}. 
The error bars for the values predicted by the ML model represent the variance of five identical models trained using different initialization seeds (five-fold cross validation). 
Overall, the ML-predicted and measured parameters exhibit qualitatively consistent trends. 
With increasing Fe($a$)/Co($b$) ratio (left to right), both measured and predicted values of  $A$ and $K_\mathrm{eff}$ decrease, $D$ remains relatively constant, and the resulting $\kappa$ increases. 
On the quantitative aspect, we observe excellent agreement for $A$. However, the measured and predicted values for $K_{\rm eff}$, $D$, and $\kappa$ exhibit varying extents of discrepancy.

\paragraph{Discrepancy Sources}
Such quantitative discrepancy may arise from predictive limitations of the model, or due to differences between the training and testing datasets used in the work.
To distinguish these sources, we tested the model on images simulated using experimentally measured sample parameters (\ref{fig:trends}: red) as inputs.
As seen in \ref{fig:trends} (blue c.f. red), all model-predicted parameters from these images exhibit excellent quantitative correspondence with the measured parameters, confirming the self-consistency of the model and reliability of the training procedure. 
This suggests that the likely source of the discrepancy is limitations inherent to the simulated dataset, which, while visually similar to experimental MFM images, exhibit several observable differences (SM \S S4).
Such differences are expected to arise from a combination of several factors unaccounted within simulations, including (a) extrinsic factors, such as polycrystalline grain structures of the sputtered films, which result in granularity of magnetic interactions, as well as defects and pinning; (b) intrinsic factors, e.g., interlayer coupling, proximity-induced effects, higher-order anisotropy etc.; as well as (c) differences in recipes between simulations and measurements. 

\paragraph{Future Training Directions}
Overall, these results establish the feasibility of ML-driven characterization of the key parameters in chiral magnetic films using microscopic images. 
The key bottleneck to full quantitative consistency between ML-predicted and experimentally measured magnetic parameters is a high-fidelity, large volume training dataset of simulated domain images sufficiently representative of experimental images. 
Future simulated datasets generated for training can incorporate granularity to mimic spatial variations in magnetic properties \cite{Legrand.2017, Juge.2019}, interlayer exchange coupling effects \cite{Chen.2023}, and field recipes \cite{Chen.2022} to better emulate experimentally measured images. % Legrand.2020
Meanwhile, ML-based techniques can serve as able complements to experimental measurements, especially in circumventing highly resource-intensive determination of magnetic parameters \cite{Boettcher.2023}. 

\section{Conclusion}

\paragraph{Results Summary}
In conclusion, we have built an ML model to estimate the key magnetic parameters governing nanoscale chiral domain phenomenology in multilayers using zero field MFM images. 
The CNN-based model was trained and optimized using micromagnetically simulated domain images, processed to remove experimentally inaccessible information. 
It achieved prediction $R^2 > 0.85$ for all parameters of interest -- $A$, $K_{\rm eff}$, $D$, and $\kappa$ -- without overfitting the data, and accurately deduced the relationship connecting $\kappa$ to the other three parameters. 
When used on MFM-measured domain images, the trained model gives reasonable predictions for the evolution of parameters across studied samples. 
Full quantitative consistency for all magnetic parameters can be achieved by enhancing the simulated training dataset to additionally incorporate extrinsic and intrinsic factors affecting imaged domain configurations, to achieve rapid all-ML characterization of chiral multilayers.

\paragraph{Impact}
Our work advances ML-based parameter prediction for multilayer films from individual determination on microscale domain configurations \cite{Kwon.2020, Kawaguchi.2021, Mamada.2021, Talapatra.2022} to  collective, or ``all-in-one'' parameter determination in  the chirality-dominated regime of functional nanoscale spin textures.
The demonstrated quantitative consistency in predicting $A$ is particularly remarkable given its large variability across used techniques, and resource-intensive needs of reliable BLS characterization \cite{Boettcher.2023}. % the exchange stiffness, 
Therefore, our built model can already be used as a valuable complement to experiments for quantifying $A$, as well as a quick, inexpensive ``first-pass'' estimation technique for all parameters.
Our workflow can be straightforwardly extended to complementary magnetic imaging techniques from the micron-scale (Kerr) to the atomic-scale (electrons, X-rays). 
With appropriate incorporation of complementary data, it can also be generalized to extract other relevant parameters of interest, such as interlayer coupling, damping, spin Hall angle etc. 
By providing the ability to unlock information stored in domain configurations, our work paves the way for sustainable, high-throughout development of ultrathin magnetic films and devices for next-generation electronics.

%%%% Methods
\section{Methods\label{sec:methods}}
\begin{small}

%\subsection{Micromagnetic simulation}
\paragraph{Micromagnetic Simulations}
\textsf{\textbf{Micromagnetic Simulations}} of a four-repeat stack were performed using the GPU-based package mumax$^3$\cite{Vansteenkiste.2014} under grain-free conditions to generate a large set of realistic domain images to train, validate, and test the CNN model.
The effective medium approximation, which treats all layers within each stack repetition as a single effective magnetic layer \cite{Woo.2016}, was used to reduce the overall simulation time. % (consisting of both magnetic and non-magnetic layers) 
Simulations were carried out on a finite-difference lattice of $1024 \times 1024  \times 4$ cells, which for a cell size of $2\times 2\times \SI{3}{\nano\meter\cubed}$, corresponds to a physical sample size of $\sim \SI{2}{\micro\meter}\times \SI{2}{\micro\meter}\times \SI{12}{\nano\meter}$. 
The ranges of simulation parameters were chosen to cover the ranges of values of experimentally accessible stacks \citep{Soumyanarayanan.2017, Chen.2022}, and are given in \ref{table:ParamRange}. 
Within these bounds, the simulation parameters were uniformly sampled using a grid.
Note that when $A$ or $K_\mathrm{eff}$ are too large, or $D$ is too small, the equilibrium domain configuration is a uniform magnetized state. 
Such parameter combinations that do not produce multi-domain magnetization configurations at ZF were excluded from the image dataset.
Finally, for each set of parameters, the magnetization image used for analysis was extracted from the second layer out of the four simulated layers after verifying that the magnetization is approximately layer-independent across the used parameters. 
  
\begin{table}
    \centering
    \begin{ruledtabular}
    \def\arraystretch{1.5} % 1 is the default
    \begin{tabular}{lccc}
    Parameter & Units & Lower Bound & Upper Bound \\ 
    \colrule
    $A$ & pJ/m & 10 & 18 \\
    $K_{\rm eff}$ & MJ/m$^3$ & 0.0 & 0.4 \\ 
    $D$ & mJ/m$^2$ & 1.2 & 2.1 \\
    \end{tabular}
    \end{ruledtabular}
    \caption{\label{table:ParamRange} Ranges of magnetic parameters,  $A$, $K_{\rm eff}$, and $D$, used for micromagnetic simulations to generate the training dataset. 
    Parameter combinations resulting in multi-domain configurations at ZF were used for training and testing.}
\end{table}

%\subsection{Image processing}
\paragraph{Method: Image processing}
\textsf{\textbf{Image Processing.}} 
Standard routines from the \textsf{CV2} and \textsf{Scikit-Image} Python libraries were used to pre-process our domain images before training the model. 
First, a median filter with a large kernel (radius: 8 pixels) was used to remove isolated salt and pepper noise. 
Next, the smoothness of the domain boundary was considered, which, for MFM-measured images, is sensitive to the imaging conditions. 
Since the smoothness is largely insensitive to the intrinsic magnetic properties, one would not expect it to measurably affect the parameter predictions. 
However, during training, the ML models were found to be sensitive to the domain boundary curvature. 
Therefore, the image-processing was used to ensure that the domain boundaries were consistently smooth across both simulation and experimental images. 
Successive median filters with a small kernel (radius: 3 pixels) were applied to gradually smoothen the domain boundaries along their length. 
For the last step, the image was filtered using a binary threshold, converting it from grayscale into black-white, eliminating any speckle noise and DW information. 
The resulting DWs are sharp, and the magnetization within a domain is effectively uniform. 
The MFM-measured images were found to be more sensitive to the choice of threshold than simulated images.
To ensure accurate and reproducible thresholding, the Otsu method \citep{Otsu.1979} was employed to determine the optimal threshold for each image ($\sim 126-127$ on the $0-255$ scale). 

%\subsection{Model architecture and training}
\paragraph{Method: CNN architecture}
\textsf{\textbf{Model Architecture \& Training.}} 
Our optimized CNN model, determined following several rounds of hyperparameter optimization, consists of 10 convolutional layers followed by 4 fully connected layers. 
All convolution filters are $3 \times 3$ in size with strides of 1. 
Max pooling is applied after the 2$^{\rm nd}$, 4$^{\rm th}$, 7$^{\rm th}$, and 10$^{\rm th}$ convolution layers, with kernels of size $2 \times 2$ and strides of 1. 
The number of convolutional filters doubles after each pooling layer. 
The subsequent 3 fully connected (dense) layers all have equal width of 256 neurons, and  are enhanced with a small dropout rate. 
The last layer consists of a fully connected layer with 4 neurons corresponding to the 4 output parameters of interest.
The last layer is the output layer, which consists of 4 neurons corresponding to the 4 magnetic parameters of interest. 
Leaky ReLu with a small slope of 0.2 was used between the fully connected layers. 
Dropout layers are included to reduce overfitting by randomly removing some nodes (and all their incoming and outgoing connections) during training \citep{Hinton.2012}. 

The model was trained for 400 epochs, with mean absolute error (MAE) as the loss function. 
9,600 images were used for training, while the remaining 2,400 images were used for validation and testing. 
A learning rate scheduler was used during the training, such that the learning rate started large and decreased gradually as the training progressed. 
This allows the ML model to learn quickly at the start, while being able to fine-tune its weights as the learning plateaus.

%\subsection{MFM imaging}
\paragraph{Method: MFM}
\textsf{\textbf{MFM Imaging}} of the multilayer thin films was performed using the Veeco Dimension\texttrademark\, 3100 scanning probe microscope. 
SSS-MFMR\texttrademark\, tips with sharp tip profile (diameter: 30\,nm) and ultra-low magnetization (Co-alloy coating, 80\,emu/cm$^3$) were used for high-resolution imaging with minimal stray field perturbations. 
All MFM images were acquired at tip-sample lift heights of 20\,nm, with an image resolution of $2048\times 2048$ pixels for $4\times$  \SI{4}{\micro\meter} scan dimensions. 
The multilayers were imaged at zero field, following \emph{ex situ} out-of-plane saturation at -400\,mT.

\end{small}

\noindent \begin{center}
{\small{}\rule[0.5ex]{0.4\columnwidth}{0.5pt}}{\small\par}
\par\end{center}

\noindent We acknowledge helpful inputs from Hang Khume Tan, Constantin C. Chirila, and Nathaniel Ng, as well as the support of the National Supercomputing Centre (NSCC), Singapore, for computational resources. 
This work was supported by the SpOT-LITE program (A*STAR Grant No. A18A6b0057), funded by Singapore's RIE2020 initiatives.

\phantomsection\addcontentsline{toc}{section}{\refname}
\bibliographystyle{apsrev4-2}
\bibliography{SkML_Refs}

\begin{thebibliography}{48}%
\makeatletter
\providecommand \@ifxundefined [1]{%
 \@ifx{#1\undefined}
}%
\providecommand \@ifnum [1]{%
 \ifnum #1\expandafter \@firstoftwo
 \else \expandafter \@secondoftwo
 \fi
}%
\providecommand \@ifx [1]{%
 \ifx #1\expandafter \@firstoftwo
 \else \expandafter \@secondoftwo
 \fi
}%
\providecommand \natexlab [1]{#1}%
\providecommand \enquote  [1]{``#1''}%
\providecommand \bibnamefont  [1]{#1}%
\providecommand \bibfnamefont [1]{#1}%
\providecommand \citenamefont [1]{#1}%
\providecommand \href@noop [0]{\@secondoftwo}%
\providecommand \href [0]{\begingroup \@sanitize@url \@href}%
\providecommand \@href[1]{\@@startlink{#1}\@@href}%
\providecommand \@@href[1]{\endgroup#1\@@endlink}%
\providecommand \@sanitize@url [0]{\catcode `\\12\catcode `\$12\catcode
  `\&12\catcode `\#12\catcode `\^12\catcode `\_12\catcode `\%12\relax}%
\providecommand \@@startlink[1]{}%
\providecommand \@@endlink[0]{}%
\providecommand \url  [0]{\begingroup\@sanitize@url \@url }%
\providecommand \@url [1]{\endgroup\@href {#1}{\urlprefix }}%
\providecommand \urlprefix  [0]{URL }%
\providecommand \Eprint [0]{\href }%
\providecommand \doibase [0]{https://doi.org/}%
\providecommand \selectlanguage [0]{\@gobble}%
\providecommand \bibinfo  [0]{\@secondoftwo}%
\providecommand \bibfield  [0]{\@secondoftwo}%
\providecommand \translation [1]{[#1]}%
\providecommand \BibitemOpen [0]{}%
\providecommand \bibitemStop [0]{}%
\providecommand \bibitemNoStop [0]{.\EOS\space}%
\providecommand \EOS [0]{\spacefactor3000\relax}%
\providecommand \BibitemShut  [1]{\csname bibitem#1\endcsname}%
\let\auto@bib@innerbib\@empty
%</preamble>
\bibitem [{\citenamefont {Nagaosa}\ and\ \citenamefont
  {Tokura}(2013)}]{Nagaosa.2013}%
  \BibitemOpen
  \bibfield  {author} {\bibinfo {author} {\bibfnamefont {N.}~\bibnamefont
  {Nagaosa}}\ and\ \bibinfo {author} {\bibfnamefont {Y.}~\bibnamefont
  {Tokura}},\ }\href {https://doi.org/10.1038/nnano.2013.243} {\bibfield
  {journal} {\bibinfo  {journal} {Nature Nanotechnology}\ }\textbf {\bibinfo
  {volume} {8}},\ \bibinfo {pages} {899} (\bibinfo {year} {2013})}\BibitemShut
  {NoStop}%
\bibitem [{\citenamefont {Soumyanarayanan}\ \emph {et~al.}(2016)\citenamefont
  {Soumyanarayanan}, \citenamefont {Reyren}, \citenamefont {Fert},\ and\
  \citenamefont {Panagopoulos}}]{Soumyanarayanan.2016}%
  \BibitemOpen
  \bibfield  {author} {\bibinfo {author} {\bibfnamefont {A.}~\bibnamefont
  {Soumyanarayanan}}, \bibinfo {author} {\bibfnamefont {N.}~\bibnamefont
  {Reyren}}, \bibinfo {author} {\bibfnamefont {A.}~\bibnamefont {Fert}},\ and\
  \bibinfo {author} {\bibfnamefont {C.}~\bibnamefont {Panagopoulos}},\ }\href
  {https://doi.org/10.1038/nature19820} {\bibfield  {journal} {\bibinfo
  {journal} {Nature}\ }\textbf {\bibinfo {volume} {539}},\ \bibinfo {pages}
  {509} (\bibinfo {year} {2016})}\BibitemShut {NoStop}%
\bibitem [{\citenamefont {Back}\ \emph {et~al.}(2020)\citenamefont {Back},
  \citenamefont {Cros}, \citenamefont {Ebert}, \citenamefont {Everschor-Sitte},
  \citenamefont {Fert}, \citenamefont {Garst}, \citenamefont {Ma},
  \citenamefont {Mankovsky}, \citenamefont {Monchesky}, \citenamefont
  {Mostovoy}, \citenamefont {Nagaosa}, \citenamefont {Parkin}, \citenamefont
  {Pfleiderer}, \citenamefont {Reyren}, \citenamefont {Rosch}, \citenamefont
  {Taguchi}, \citenamefont {Tokura}, \citenamefont {Bergmann},\ and\
  \citenamefont {Zang}}]{Back.2020}%
  \BibitemOpen
  \bibfield  {author} {\bibinfo {author} {\bibfnamefont {C.}~\bibnamefont
  {Back}}, \bibinfo {author} {\bibfnamefont {V.}~\bibnamefont {Cros}}, \bibinfo
  {author} {\bibfnamefont {H.}~\bibnamefont {Ebert}}, \bibinfo {author}
  {\bibfnamefont {K.}~\bibnamefont {Everschor-Sitte}}, \bibinfo {author}
  {\bibfnamefont {A.}~\bibnamefont {Fert}}, \bibinfo {author} {\bibfnamefont
  {M.}~\bibnamefont {Garst}}, \bibinfo {author} {\bibfnamefont
  {T.}~\bibnamefont {Ma}}, \bibinfo {author} {\bibfnamefont {S.}~\bibnamefont
  {Mankovsky}}, \bibinfo {author} {\bibfnamefont {T.~L.}\ \bibnamefont
  {Monchesky}}, \bibinfo {author} {\bibfnamefont {M.}~\bibnamefont {Mostovoy}},
  \bibinfo {author} {\bibfnamefont {N.}~\bibnamefont {Nagaosa}}, \bibinfo
  {author} {\bibfnamefont {S.~S.~P.}\ \bibnamefont {Parkin}}, \bibinfo {author}
  {\bibfnamefont {C.}~\bibnamefont {Pfleiderer}}, \bibinfo {author}
  {\bibfnamefont {N.}~\bibnamefont {Reyren}}, \bibinfo {author} {\bibfnamefont
  {A.}~\bibnamefont {Rosch}}, \bibinfo {author} {\bibfnamefont
  {Y.}~\bibnamefont {Taguchi}}, \bibinfo {author} {\bibfnamefont
  {Y.}~\bibnamefont {Tokura}}, \bibinfo {author} {\bibfnamefont {K.~v.}\
  \bibnamefont {Bergmann}},\ and\ \bibinfo {author} {\bibfnamefont
  {J.}~\bibnamefont {Zang}},\ }\href {https://doi.org/10.1088/1361-6463/ab8418}
  {\bibfield  {journal} {\bibinfo  {journal} {Journal of Physics D: Applied
  Physics}\ }\textbf {\bibinfo {volume} {53}},\ \bibinfo {pages} {363001}
  (\bibinfo {year} {2020})}\BibitemShut {NoStop}%
\bibitem [{\citenamefont {G\"{o}bel}\ \emph {et~al.}(2021)\citenamefont
  {G\"{o}bel}, \citenamefont {Mertig},\ and\ \citenamefont
  {Tretiakov}}]{Goebel.2021}%
  \BibitemOpen
  \bibfield  {author} {\bibinfo {author} {\bibfnamefont {B.}~\bibnamefont
  {G\"{o}bel}}, \bibinfo {author} {\bibfnamefont {I.}~\bibnamefont {Mertig}},\
  and\ \bibinfo {author} {\bibfnamefont {O.~A.}\ \bibnamefont {Tretiakov}},\
  }\href {https://doi.org/10.1016/j.physrep.2020.10.001} {\bibfield  {journal}
  {\bibinfo  {journal} {Physics Reports}\ }\textbf {\bibinfo {volume} {895}},\
  \bibinfo {pages} {1} (\bibinfo {year} {2021})}\BibitemShut {NoStop}%
\bibitem [{\citenamefont {Fert}\ \emph {et~al.}(2017)\citenamefont {Fert},
  \citenamefont {Reyren},\ and\ \citenamefont {Cros}}]{Fert.2017}%
  \BibitemOpen
  \bibfield  {author} {\bibinfo {author} {\bibfnamefont {A.}~\bibnamefont
  {Fert}}, \bibinfo {author} {\bibfnamefont {N.}~\bibnamefont {Reyren}},\ and\
  \bibinfo {author} {\bibfnamefont {V.}~\bibnamefont {Cros}},\ }\href
  {https://doi.org/10.1038/natrevmats.2017.31} {\bibfield  {journal} {\bibinfo
  {journal} {Nature Reviews Materials}\ }\textbf {\bibinfo {volume} {2}},\
  \bibinfo {pages} {17031} (\bibinfo {year} {2017})}\BibitemShut {NoStop}%
\bibitem [{\citenamefont {Song}\ \emph {et~al.}(2020)\citenamefont {Song},
  \citenamefont {Jeong}, \citenamefont {Pan}, \citenamefont {Zhang},
  \citenamefont {Xia}, \citenamefont {Cha}, \citenamefont {Park}, \citenamefont
  {Kim}, \citenamefont {Finizio}, \citenamefont {Raabe}, \citenamefont {Chang},
  \citenamefont {Zhou}, \citenamefont {Zhao}, \citenamefont {Kang},
  \citenamefont {Ju},\ and\ \citenamefont {Woo}}]{Song.2020}%
  \BibitemOpen
  \bibfield  {author} {\bibinfo {author} {\bibfnamefont {K.~M.}\ \bibnamefont
  {Song}}, \bibinfo {author} {\bibfnamefont {J.-S.}\ \bibnamefont {Jeong}},
  \bibinfo {author} {\bibfnamefont {B.}~\bibnamefont {Pan}}, \bibinfo {author}
  {\bibfnamefont {X.}~\bibnamefont {Zhang}}, \bibinfo {author} {\bibfnamefont
  {J.}~\bibnamefont {Xia}}, \bibinfo {author} {\bibfnamefont {S.}~\bibnamefont
  {Cha}}, \bibinfo {author} {\bibfnamefont {T.-E.}\ \bibnamefont {Park}},
  \bibinfo {author} {\bibfnamefont {K.}~\bibnamefont {Kim}}, \bibinfo {author}
  {\bibfnamefont {S.}~\bibnamefont {Finizio}}, \bibinfo {author} {\bibfnamefont
  {J.}~\bibnamefont {Raabe}}, \bibinfo {author} {\bibfnamefont
  {J.}~\bibnamefont {Chang}}, \bibinfo {author} {\bibfnamefont
  {Y.}~\bibnamefont {Zhou}}, \bibinfo {author} {\bibfnamefont {W.}~\bibnamefont
  {Zhao}}, \bibinfo {author} {\bibfnamefont {W.}~\bibnamefont {Kang}}, \bibinfo
  {author} {\bibfnamefont {H.}~\bibnamefont {Ju}},\ and\ \bibinfo {author}
  {\bibfnamefont {S.}~\bibnamefont {Woo}},\ }\href
  {https://doi.org/10.1038/s41928-020-0385-0} {\bibfield  {journal} {\bibinfo
  {journal} {Nature Electronics}\ }\textbf {\bibinfo {volume} {3}},\ \bibinfo
  {pages} {148} (\bibinfo {year} {2020})}\BibitemShut {NoStop}%
\bibitem [{\citenamefont {Bourianoff}\ \emph {et~al.}(2018)\citenamefont
  {Bourianoff}, \citenamefont {Pinna}, \citenamefont {Sitte},\ and\
  \citenamefont {Everschor-Sitte}}]{Bourianoff.2018}%
  \BibitemOpen
  \bibfield  {author} {\bibinfo {author} {\bibfnamefont {G.}~\bibnamefont
  {Bourianoff}}, \bibinfo {author} {\bibfnamefont {D.}~\bibnamefont {Pinna}},
  \bibinfo {author} {\bibfnamefont {M.}~\bibnamefont {Sitte}},\ and\ \bibinfo
  {author} {\bibfnamefont {K.}~\bibnamefont {Everschor-Sitte}},\ }\href
  {https://doi.org/10.1063/1.5006918} {\bibfield  {journal} {\bibinfo
  {journal} {AIP Advances}\ }\textbf {\bibinfo {volume} {8}},\ \bibinfo {pages}
  {055602} (\bibinfo {year} {2018})}\BibitemShut {NoStop}%
\bibitem [{\citenamefont {Pinna}\ \emph {et~al.}(2018)\citenamefont {Pinna},
  \citenamefont {Araujo}, \citenamefont {Kim}, \citenamefont {Cros},
  \citenamefont {Querlioz}, \citenamefont {Bessiere}, \citenamefont {Droulez},\
  and\ \citenamefont {Grollier}}]{Pinna.2018}%
  \BibitemOpen
  \bibfield  {author} {\bibinfo {author} {\bibfnamefont {D.}~\bibnamefont
  {Pinna}}, \bibinfo {author} {\bibfnamefont {F.~A.}\ \bibnamefont {Araujo}},
  \bibinfo {author} {\bibfnamefont {J.-V.}\ \bibnamefont {Kim}}, \bibinfo
  {author} {\bibfnamefont {V.}~\bibnamefont {Cros}}, \bibinfo {author}
  {\bibfnamefont {D.}~\bibnamefont {Querlioz}}, \bibinfo {author}
  {\bibfnamefont {P.}~\bibnamefont {Bessiere}}, \bibinfo {author}
  {\bibfnamefont {J.}~\bibnamefont {Droulez}},\ and\ \bibinfo {author}
  {\bibfnamefont {J.}~\bibnamefont {Grollier}},\ }\href
  {https://doi.org/10.1103/physrevapplied.9.064018} {\bibfield  {journal}
  {\bibinfo  {journal} {Physical Review Applied}\ }\textbf {\bibinfo {volume}
  {9}},\ \bibinfo {pages} {064018} (\bibinfo {year} {2018})}\BibitemShut
  {NoStop}%
\bibitem [{\citenamefont {Bogdanov}\ and\ \citenamefont
  {Rößler}(2001)}]{Bogdanov.2001}%
  \BibitemOpen
  \bibfield  {author} {\bibinfo {author} {\bibfnamefont {A.~N.}\ \bibnamefont
  {Bogdanov}}\ and\ \bibinfo {author} {\bibfnamefont {U.~K.}\ \bibnamefont
  {Rößler}},\ }\href {https://doi.org/10.1103/physrevlett.87.037203}
  {\bibfield  {journal} {\bibinfo  {journal} {Physical Review Letters}\
  }\textbf {\bibinfo {volume} {87}},\ \bibinfo {pages} {037203} (\bibinfo
  {year} {2001})}\BibitemShut {NoStop}%
\bibitem [{\citenamefont {Woo}\ \emph {et~al.}(2016)\citenamefont {Woo},
  \citenamefont {Litzius}, \citenamefont {Krüger}, \citenamefont {Im},
  \citenamefont {Caretta}, \citenamefont {Richter}, \citenamefont {Mann},
  \citenamefont {Krone}, \citenamefont {Reeve}, \citenamefont {Weigand},
  \citenamefont {Agrawal}, \citenamefont {Lemesh}, \citenamefont {Mawass},
  \citenamefont {Fischer}, \citenamefont {Kläui},\ and\ \citenamefont
  {Beach}}]{Woo.2016}%
  \BibitemOpen
  \bibfield  {author} {\bibinfo {author} {\bibfnamefont {S.}~\bibnamefont
  {Woo}}, \bibinfo {author} {\bibfnamefont {K.}~\bibnamefont {Litzius}},
  \bibinfo {author} {\bibfnamefont {B.}~\bibnamefont {Krüger}}, \bibinfo
  {author} {\bibfnamefont {M.-Y.}\ \bibnamefont {Im}}, \bibinfo {author}
  {\bibfnamefont {L.}~\bibnamefont {Caretta}}, \bibinfo {author} {\bibfnamefont
  {K.}~\bibnamefont {Richter}}, \bibinfo {author} {\bibfnamefont
  {M.}~\bibnamefont {Mann}}, \bibinfo {author} {\bibfnamefont {A.}~\bibnamefont
  {Krone}}, \bibinfo {author} {\bibfnamefont {R.~M.}\ \bibnamefont {Reeve}},
  \bibinfo {author} {\bibfnamefont {M.}~\bibnamefont {Weigand}}, \bibinfo
  {author} {\bibfnamefont {P.}~\bibnamefont {Agrawal}}, \bibinfo {author}
  {\bibfnamefont {I.}~\bibnamefont {Lemesh}}, \bibinfo {author} {\bibfnamefont
  {M.-A.}\ \bibnamefont {Mawass}}, \bibinfo {author} {\bibfnamefont
  {P.}~\bibnamefont {Fischer}}, \bibinfo {author} {\bibfnamefont
  {M.}~\bibnamefont {Kläui}},\ and\ \bibinfo {author} {\bibfnamefont
  {G.~S.~D.}\ \bibnamefont {Beach}},\ }\href {https://doi.org/10.1038/nmat4593}
  {\bibfield  {journal} {\bibinfo  {journal} {Nature Materials}\ }\textbf
  {\bibinfo {volume} {15}},\ \bibinfo {pages} {501} (\bibinfo {year}
  {2016})}\BibitemShut {NoStop}%
\bibitem [{\citenamefont {Soumyanarayanan}\ \emph {et~al.}(2017)\citenamefont
  {Soumyanarayanan}, \citenamefont {Raju}, \citenamefont {Oyarce},
  \citenamefont {Tan}, \citenamefont {Im}, \citenamefont {Petrović},
  \citenamefont {Ho}, \citenamefont {Khoo}, \citenamefont {Tran}, \citenamefont
  {Gan}, \citenamefont {Ernult},\ and\ \citenamefont
  {Panagopoulos}}]{Soumyanarayanan.2017}%
  \BibitemOpen
  \bibfield  {author} {\bibinfo {author} {\bibfnamefont {A.}~\bibnamefont
  {Soumyanarayanan}}, \bibinfo {author} {\bibfnamefont {M.}~\bibnamefont
  {Raju}}, \bibinfo {author} {\bibfnamefont {A.~L.~G.}\ \bibnamefont {Oyarce}},
  \bibinfo {author} {\bibfnamefont {A.~K.~C.}\ \bibnamefont {Tan}}, \bibinfo
  {author} {\bibfnamefont {M.-Y.}\ \bibnamefont {Im}}, \bibinfo {author}
  {\bibfnamefont {A.~P.}\ \bibnamefont {Petrović}}, \bibinfo {author}
  {\bibfnamefont {P.}~\bibnamefont {Ho}}, \bibinfo {author} {\bibfnamefont
  {K.~H.}\ \bibnamefont {Khoo}}, \bibinfo {author} {\bibfnamefont
  {M.}~\bibnamefont {Tran}}, \bibinfo {author} {\bibfnamefont {C.~K.}\
  \bibnamefont {Gan}}, \bibinfo {author} {\bibfnamefont {F.}~\bibnamefont
  {Ernult}},\ and\ \bibinfo {author} {\bibfnamefont {C.}~\bibnamefont
  {Panagopoulos}},\ }\href {https://doi.org/10.1038/nmat4934} {\bibfield
  {journal} {\bibinfo  {journal} {Nature Materials}\ }\textbf {\bibinfo
  {volume} {16}},\ \bibinfo {pages} {898} (\bibinfo {year} {2017})}\BibitemShut
  {NoStop}%
\bibitem [{\citenamefont {Johnson}\ \emph {et~al.}(1996)\citenamefont
  {Johnson}, \citenamefont {Bloemen}, \citenamefont {den Broeder},\ and\
  \citenamefont {de~Vries}}]{Johnson.1996}%
  \BibitemOpen
  \bibfield  {author} {\bibinfo {author} {\bibfnamefont {M.~T.}\ \bibnamefont
  {Johnson}}, \bibinfo {author} {\bibfnamefont {P.~J.~H.}\ \bibnamefont
  {Bloemen}}, \bibinfo {author} {\bibfnamefont {F.~J.~A.}\ \bibnamefont {den
  Broeder}},\ and\ \bibinfo {author} {\bibfnamefont {J.~J.}\ \bibnamefont
  {de~Vries}},\ }\href {https://doi.org/10.1088/0034-4885/59/11/002} {\bibfield
   {journal} {\bibinfo  {journal} {Reports on Progress in Physics}\ }\textbf
  {\bibinfo {volume} {59}},\ \bibinfo {pages} {1409} (\bibinfo {year}
  {1996})}\BibitemShut {NoStop}%
\bibitem [{\citenamefont {Kuepferling}\ \emph {et~al.}(2020)\citenamefont
  {Kuepferling}, \citenamefont {Casiraghi}, \citenamefont {Soares},
  \citenamefont {Durin}, \citenamefont {Garcia-Sanchez}, \citenamefont {Chen},
  \citenamefont {Back}, \citenamefont {Marrows}, \citenamefont {Tacchi},\ and\
  \citenamefont {Carlotti}}]{Kuepferling.2020}%
  \BibitemOpen
  \bibfield  {author} {\bibinfo {author} {\bibfnamefont {M.}~\bibnamefont
  {Kuepferling}}, \bibinfo {author} {\bibfnamefont {A.}~\bibnamefont
  {Casiraghi}}, \bibinfo {author} {\bibfnamefont {G.}~\bibnamefont {Soares}},
  \bibinfo {author} {\bibfnamefont {G.}~\bibnamefont {Durin}}, \bibinfo
  {author} {\bibfnamefont {F.}~\bibnamefont {Garcia-Sanchez}}, \bibinfo
  {author} {\bibfnamefont {L.}~\bibnamefont {Chen}}, \bibinfo {author}
  {\bibfnamefont {C.~H.}\ \bibnamefont {Back}}, \bibinfo {author}
  {\bibfnamefont {C.~H.}\ \bibnamefont {Marrows}}, \bibinfo {author}
  {\bibfnamefont {S.}~\bibnamefont {Tacchi}},\ and\ \bibinfo {author}
  {\bibfnamefont {G.}~\bibnamefont {Carlotti}},\ }\href@noop {} {\bibfield
  {journal} {\bibinfo  {journal} {ArXiv E-Prints}\ } (\bibinfo {year}
  {2020})},\ \Eprint {https://arxiv.org/abs/2009.11830} {2009.11830}
  \BibitemShut {NoStop}%
\bibitem [{\citenamefont {B\"{o}ttcher}\ \emph
  {et~al.}(2021{\natexlab{a}})\citenamefont {B\"{o}ttcher}, \citenamefont
  {Lee}, \citenamefont {Heussner}, \citenamefont {Jaiswal}, \citenamefont
  {Jakob}, \citenamefont {Kläui}, \citenamefont {Hillebrands}, \citenamefont
  {Brächer},\ and\ \citenamefont {Pirro}}]{Boettcher.2021}%
  \BibitemOpen
  \bibfield  {author} {\bibinfo {author} {\bibfnamefont {T.}~\bibnamefont
  {B\"{o}ttcher}}, \bibinfo {author} {\bibfnamefont {K.}~\bibnamefont {Lee}},
  \bibinfo {author} {\bibfnamefont {F.}~\bibnamefont {Heussner}}, \bibinfo
  {author} {\bibfnamefont {S.}~\bibnamefont {Jaiswal}}, \bibinfo {author}
  {\bibfnamefont {G.}~\bibnamefont {Jakob}}, \bibinfo {author} {\bibfnamefont
  {M.}~\bibnamefont {Kläui}}, \bibinfo {author} {\bibfnamefont
  {B.}~\bibnamefont {Hillebrands}}, \bibinfo {author} {\bibfnamefont
  {T.}~\bibnamefont {Brächer}},\ and\ \bibinfo {author} {\bibfnamefont
  {P.}~\bibnamefont {Pirro}},\ }\href
  {https://doi.org/10.1109/tmag.2021.3079259} {\bibfield  {journal} {\bibinfo
  {journal} {IEEE Transactions on Magnetics}\ }\textbf {\bibinfo {volume}
  {57}},\ \bibinfo {pages} {1600207} (\bibinfo {year}
  {2021}{\natexlab{a}})}\BibitemShut {NoStop}%
\bibitem [{\citenamefont {B\"{o}ttcher}\ \emph
  {et~al.}(2021{\natexlab{b}})\citenamefont {B\"{o}ttcher}, \citenamefont
  {Suraj}, \citenamefont {Chen}, \citenamefont {Sinha}, \citenamefont {Tan},
  \citenamefont {Tan}, \citenamefont {Hillebrands}, \citenamefont {Kostylev},
  \citenamefont {Laskowski}, \citenamefont {Khoo}, \citenamefont
  {Soumyanarayanan},\ and\ \citenamefont {Pirro}}]{Boettcher.2023}%
  \BibitemOpen
  \bibfield  {author} {\bibinfo {author} {\bibfnamefont {T.}~\bibnamefont
  {B\"{o}ttcher}}, \bibinfo {author} {\bibfnamefont {T.}~\bibnamefont {Suraj}},
  \bibinfo {author} {\bibfnamefont {X.}~\bibnamefont {Chen}}, \bibinfo {author}
  {\bibfnamefont {B.}~\bibnamefont {Sinha}}, \bibinfo {author} {\bibfnamefont
  {H.~R.}\ \bibnamefont {Tan}}, \bibinfo {author} {\bibfnamefont {H.~K.}\
  \bibnamefont {Tan}}, \bibinfo {author} {\bibfnamefont {B.}~\bibnamefont
  {Hillebrands}}, \bibinfo {author} {\bibfnamefont {M.}~\bibnamefont
  {Kostylev}}, \bibinfo {author} {\bibfnamefont {R.}~\bibnamefont {Laskowski}},
  \bibinfo {author} {\bibfnamefont {K.~H.}\ \bibnamefont {Khoo}}, \bibinfo
  {author} {\bibfnamefont {A.}~\bibnamefont {Soumyanarayanan}},\ and\ \bibinfo
  {author} {\bibfnamefont {P.}~\bibnamefont {Pirro}},\ }\href
  {https://doi.org/10.1103/physrevb.107.094405} {\bibfield  {journal} {\bibinfo
   {journal} {Physical Review B}\ }\textbf {\bibinfo {volume} {107}},\ \bibinfo
  {pages} {094405} (\bibinfo {year} {2021}{\natexlab{b}})}\BibitemShut
  {NoStop}%
\bibitem [{\citenamefont {Je}\ \emph {et~al.}(2013)\citenamefont {Je},
  \citenamefont {Kim}, \citenamefont {Yoo}, \citenamefont {Min}, \citenamefont
  {Lee},\ and\ \citenamefont {Choe}}]{Je.2013}%
  \BibitemOpen
  \bibfield  {author} {\bibinfo {author} {\bibfnamefont {S.-G.}\ \bibnamefont
  {Je}}, \bibinfo {author} {\bibfnamefont {D.-H.}\ \bibnamefont {Kim}},
  \bibinfo {author} {\bibfnamefont {S.-C.}\ \bibnamefont {Yoo}}, \bibinfo
  {author} {\bibfnamefont {B.-C.}\ \bibnamefont {Min}}, \bibinfo {author}
  {\bibfnamefont {K.-J.}\ \bibnamefont {Lee}},\ and\ \bibinfo {author}
  {\bibfnamefont {S.-B.}\ \bibnamefont {Choe}},\ }\href
  {https://doi.org/10.1103/physrevb.88.214401} {\bibfield  {journal} {\bibinfo
  {journal} {Physical Review B}\ }\textbf {\bibinfo {volume} {88}},\ \bibinfo
  {pages} {214401} (\bibinfo {year} {2013})}\BibitemShut {NoStop}%
\bibitem [{\citenamefont {Hrabec}\ \emph {et~al.}(2014)\citenamefont {Hrabec},
  \citenamefont {Porter}, \citenamefont {Wells}, \citenamefont {Benitez},
  \citenamefont {Burnell}, \citenamefont {McVitie}, \citenamefont {McGrouther},
  \citenamefont {Moore},\ and\ \citenamefont {Marrows}}]{Hrabec.2014}%
  \BibitemOpen
  \bibfield  {author} {\bibinfo {author} {\bibfnamefont {A.}~\bibnamefont
  {Hrabec}}, \bibinfo {author} {\bibfnamefont {N.~A.}\ \bibnamefont {Porter}},
  \bibinfo {author} {\bibfnamefont {A.}~\bibnamefont {Wells}}, \bibinfo
  {author} {\bibfnamefont {M.~J.}\ \bibnamefont {Benitez}}, \bibinfo {author}
  {\bibfnamefont {G.}~\bibnamefont {Burnell}}, \bibinfo {author} {\bibfnamefont
  {S.}~\bibnamefont {McVitie}}, \bibinfo {author} {\bibfnamefont
  {D.}~\bibnamefont {McGrouther}}, \bibinfo {author} {\bibfnamefont {T.~A.}\
  \bibnamefont {Moore}},\ and\ \bibinfo {author} {\bibfnamefont {C.~H.}\
  \bibnamefont {Marrows}},\ }\href {https://doi.org/10.1103/physrevb.90.020402}
  {\bibfield  {journal} {\bibinfo  {journal} {Physical Review B}\ }\textbf
  {\bibinfo {volume} {90}},\ \bibinfo {pages} {020402} (\bibinfo {year}
  {2014})}\BibitemShut {NoStop}%
\bibitem [{\citenamefont {Vaz}\ \emph {et~al.}(2008)\citenamefont {Vaz},
  \citenamefont {Bland},\ and\ \citenamefont {Lauhoff}}]{Vaz.2008}%
  \BibitemOpen
  \bibfield  {author} {\bibinfo {author} {\bibfnamefont {C.~A.~F.}\
  \bibnamefont {Vaz}}, \bibinfo {author} {\bibfnamefont {J.~A.~C.}\
  \bibnamefont {Bland}},\ and\ \bibinfo {author} {\bibfnamefont
  {G.}~\bibnamefont {Lauhoff}},\ }\href
  {https://doi.org/10.1088/0034-4885/71/5/056501} {\bibfield  {journal}
  {\bibinfo  {journal} {Reports on Progress in Physics}\ }\textbf {\bibinfo
  {volume} {71}},\ \bibinfo {pages} {056501} (\bibinfo {year}
  {2008})}\BibitemShut {NoStop}%
\bibitem [{\citenamefont {Nembach}\ \emph {et~al.}(2015)\citenamefont
  {Nembach}, \citenamefont {Shaw}, \citenamefont {Weiler}, \citenamefont
  {Jué},\ and\ \citenamefont {Silva}}]{Nembach.2015}%
  \BibitemOpen
  \bibfield  {author} {\bibinfo {author} {\bibfnamefont {H.~T.}\ \bibnamefont
  {Nembach}}, \bibinfo {author} {\bibfnamefont {J.~M.}\ \bibnamefont {Shaw}},
  \bibinfo {author} {\bibfnamefont {M.}~\bibnamefont {Weiler}}, \bibinfo
  {author} {\bibfnamefont {E.}~\bibnamefont {Jué}},\ and\ \bibinfo {author}
  {\bibfnamefont {T.~J.}\ \bibnamefont {Silva}},\ }\href
  {https://doi.org/10.1038/nphys3418} {\bibfield  {journal} {\bibinfo
  {journal} {Nature Physics}\ }\textbf {\bibinfo {volume} {11}},\ \bibinfo
  {pages} {825} (\bibinfo {year} {2015})}\BibitemShut {NoStop}%
\bibitem [{\citenamefont {Donahue}(1998)}]{Donahue.1998}%
  \BibitemOpen
  \bibfield  {author} {\bibinfo {author} {\bibfnamefont {M.~J.}\ \bibnamefont
  {Donahue}},\ }\href {https://doi.org/10.1063/1.367690} {\bibfield  {journal}
  {\bibinfo  {journal} {Journal of Applied Physics}\ }\textbf {\bibinfo
  {volume} {4}},\ \bibinfo {pages} {6491} (\bibinfo {year} {1998})}\BibitemShut
  {NoStop}%
\bibitem [{\citenamefont {Vansteenkiste}\ \emph {et~al.}(2014)\citenamefont
  {Vansteenkiste}, \citenamefont {Leliaert}, \citenamefont {Dvornik},
  \citenamefont {Helsen}, \citenamefont {Garcia-Sanchez},\ and\ \citenamefont
  {Van~Waeyenberge}}]{Vansteenkiste.2014}%
  \BibitemOpen
  \bibfield  {author} {\bibinfo {author} {\bibfnamefont {A.}~\bibnamefont
  {Vansteenkiste}}, \bibinfo {author} {\bibfnamefont {J.}~\bibnamefont
  {Leliaert}}, \bibinfo {author} {\bibfnamefont {M.}~\bibnamefont {Dvornik}},
  \bibinfo {author} {\bibfnamefont {M.}~\bibnamefont {Helsen}}, \bibinfo
  {author} {\bibfnamefont {F.}~\bibnamefont {Garcia-Sanchez}},\ and\ \bibinfo
  {author} {\bibfnamefont {B.}~\bibnamefont {Van~Waeyenberge}},\ }\href
  {https://doi.org/10.1063/1.4899186} {\bibfield  {journal} {\bibinfo
  {journal} {AIP Advances}\ }\textbf {\bibinfo {volume} {4}},\ \bibinfo {pages}
  {107133} (\bibinfo {year} {2014})}\BibitemShut {NoStop}%
\bibitem [{\citenamefont {Moreau-Luchaire}\ \emph {et~al.}(2016)\citenamefont
  {Moreau-Luchaire}, \citenamefont {Moutafis}, \citenamefont {Reyren},
  \citenamefont {Sampaio}, \citenamefont {Vaz}, \citenamefont {Horne},
  \citenamefont {Bouzehouane}, \citenamefont {Garcia}, \citenamefont
  {Deranlot}, \citenamefont {Warnicke}, \citenamefont {Wohlhüter},
  \citenamefont {George}, \citenamefont {Weigand}, \citenamefont {Raabe},
  \citenamefont {Cros},\ and\ \citenamefont {Fert}}]{MoreauLuchaire.2016}%
  \BibitemOpen
  \bibfield  {author} {\bibinfo {author} {\bibfnamefont {C.}~\bibnamefont
  {Moreau-Luchaire}}, \bibinfo {author} {\bibfnamefont {C.}~\bibnamefont
  {Moutafis}}, \bibinfo {author} {\bibfnamefont {N.}~\bibnamefont {Reyren}},
  \bibinfo {author} {\bibfnamefont {J.}~\bibnamefont {Sampaio}}, \bibinfo
  {author} {\bibfnamefont {C.~A.~F.}\ \bibnamefont {Vaz}}, \bibinfo {author}
  {\bibfnamefont {N.~V.}\ \bibnamefont {Horne}}, \bibinfo {author}
  {\bibfnamefont {K.}~\bibnamefont {Bouzehouane}}, \bibinfo {author}
  {\bibfnamefont {K.}~\bibnamefont {Garcia}}, \bibinfo {author} {\bibfnamefont
  {C.}~\bibnamefont {Deranlot}}, \bibinfo {author} {\bibfnamefont
  {P.}~\bibnamefont {Warnicke}}, \bibinfo {author} {\bibfnamefont
  {P.}~\bibnamefont {Wohlhüter}}, \bibinfo {author} {\bibfnamefont {J.-M.}\
  \bibnamefont {George}}, \bibinfo {author} {\bibfnamefont {M.}~\bibnamefont
  {Weigand}}, \bibinfo {author} {\bibfnamefont {J.}~\bibnamefont {Raabe}},
  \bibinfo {author} {\bibfnamefont {V.}~\bibnamefont {Cros}},\ and\ \bibinfo
  {author} {\bibfnamefont {A.}~\bibnamefont {Fert}},\ }\href
  {https://doi.org/10.1038/nnano.2015.313} {\bibfield  {journal} {\bibinfo
  {journal} {Nature Nanotechnology}\ }\textbf {\bibinfo {volume} {11}},\
  \bibinfo {pages} {444} (\bibinfo {year} {2016})}\BibitemShut {NoStop}%
\bibitem [{\citenamefont {Kozlov}\ \emph {et~al.}(2020)\citenamefont {Kozlov},
  \citenamefont {Kolesnikov}, \citenamefont {Stebliy}, \citenamefont
  {Golikov},\ and\ \citenamefont {Davydenko}}]{Kozlov.2020}%
  \BibitemOpen
  \bibfield  {author} {\bibinfo {author} {\bibfnamefont {A.~G.}\ \bibnamefont
  {Kozlov}}, \bibinfo {author} {\bibfnamefont {A.~G.}\ \bibnamefont
  {Kolesnikov}}, \bibinfo {author} {\bibfnamefont {M.~E.}\ \bibnamefont
  {Stebliy}}, \bibinfo {author} {\bibfnamefont {A.~P.}\ \bibnamefont
  {Golikov}},\ and\ \bibinfo {author} {\bibfnamefont {A.~V.}\ \bibnamefont
  {Davydenko}},\ }\href {https://doi.org/10.1103/PhysRevB.102.144411}
  {\bibfield  {journal} {\bibinfo  {journal} {Physical Review B}\ }\textbf
  {\bibinfo {volume} {102}},\ \bibinfo {pages} {144411} (\bibinfo {year}
  {2020})}\BibitemShut {NoStop}%
\bibitem [{\citenamefont {Chen}\ \emph
  {et~al.}(2022{\natexlab{a}})\citenamefont {Chen}, \citenamefont {Lin},
  \citenamefont {Kong}, \citenamefont {Tan}, \citenamefont {Tan}, \citenamefont
  {Je}, \citenamefont {Tan}, \citenamefont {Khoo}, \citenamefont {Im},\ and\
  \citenamefont {Soumyanarayanan}}]{Chen.2022}%
  \BibitemOpen
  \bibfield  {author} {\bibinfo {author} {\bibfnamefont {X.}~\bibnamefont
  {Chen}}, \bibinfo {author} {\bibfnamefont {M.}~\bibnamefont {Lin}}, \bibinfo
  {author} {\bibfnamefont {J.~F.}\ \bibnamefont {Kong}}, \bibinfo {author}
  {\bibfnamefont {H.~R.}\ \bibnamefont {Tan}}, \bibinfo {author} {\bibfnamefont
  {A.~K.}\ \bibnamefont {Tan}}, \bibinfo {author} {\bibfnamefont {S.-G.}\
  \bibnamefont {Je}}, \bibinfo {author} {\bibfnamefont {H.~K.}\ \bibnamefont
  {Tan}}, \bibinfo {author} {\bibfnamefont {K.~H.}\ \bibnamefont {Khoo}},
  \bibinfo {author} {\bibfnamefont {M.-Y.}\ \bibnamefont {Im}},\ and\ \bibinfo
  {author} {\bibfnamefont {A.}~\bibnamefont {Soumyanarayanan}},\ }\href
  {https://doi.org/10.1002/advs.202103978} {\bibfield  {journal} {\bibinfo
  {journal} {Advanced Science}\ }\textbf {\bibinfo {volume} {9}},\ \bibinfo
  {pages} {2103978} (\bibinfo {year} {2022}{\natexlab{a}})}\BibitemShut
  {NoStop}%
\bibitem [{\citenamefont {Carleo}\ \emph {et~al.}(2019)\citenamefont {Carleo},
  \citenamefont {Cirac}, \citenamefont {Cranmer}, \citenamefont {Daudet},
  \citenamefont {Schuld}, \citenamefont {Tishby}, \citenamefont
  {Vogt-Maranto},\ and\ \citenamefont {Zdeborová}}]{Carleo.2019}%
  \BibitemOpen
  \bibfield  {author} {\bibinfo {author} {\bibfnamefont {G.}~\bibnamefont
  {Carleo}}, \bibinfo {author} {\bibfnamefont {I.}~\bibnamefont {Cirac}},
  \bibinfo {author} {\bibfnamefont {K.}~\bibnamefont {Cranmer}}, \bibinfo
  {author} {\bibfnamefont {L.}~\bibnamefont {Daudet}}, \bibinfo {author}
  {\bibfnamefont {M.}~\bibnamefont {Schuld}}, \bibinfo {author} {\bibfnamefont
  {N.}~\bibnamefont {Tishby}}, \bibinfo {author} {\bibfnamefont
  {L.}~\bibnamefont {Vogt-Maranto}},\ and\ \bibinfo {author} {\bibfnamefont
  {L.}~\bibnamefont {Zdeborová}},\ }\href
  {https://doi.org/10.1103/RevModPhys.91.045002} {\bibfield  {journal}
  {\bibinfo  {journal} {Reviews of Modern Physics}\ }\textbf {\bibinfo {volume}
  {91}},\ \bibinfo {pages} {045002} (\bibinfo {year} {2019})}\BibitemShut
  {NoStop}%
\bibitem [{\citenamefont {Schmidt}\ \emph {et~al.}(2019)\citenamefont
  {Schmidt}, \citenamefont {Marques}, \citenamefont {Botti},\ and\
  \citenamefont {Marques}}]{Schmidt.2019}%
  \BibitemOpen
  \bibfield  {author} {\bibinfo {author} {\bibfnamefont {J.}~\bibnamefont
  {Schmidt}}, \bibinfo {author} {\bibfnamefont {M.~R.~G.}\ \bibnamefont
  {Marques}}, \bibinfo {author} {\bibfnamefont {S.}~\bibnamefont {Botti}},\
  and\ \bibinfo {author} {\bibfnamefont {M.~A.~L.}\ \bibnamefont {Marques}},\
  }\href {https://doi.org/10.1038/s41524-019-0221-0} {\bibfield  {journal}
  {\bibinfo  {journal} {npj Computational Materials}\ }\textbf {\bibinfo
  {volume} {5}},\ \bibinfo {pages} {1} (\bibinfo {year} {2019})}\BibitemShut
  {NoStop}%
\bibitem [{\citenamefont {Iakovlev}\ \emph {et~al.}(2018)\citenamefont
  {Iakovlev}, \citenamefont {Sotnikov},\ and\ \citenamefont
  {Mazurenko}}]{Iakovlev.2018}%
  \BibitemOpen
  \bibfield  {author} {\bibinfo {author} {\bibfnamefont {I.~A.}\ \bibnamefont
  {Iakovlev}}, \bibinfo {author} {\bibfnamefont {O.~M.}\ \bibnamefont
  {Sotnikov}},\ and\ \bibinfo {author} {\bibfnamefont {V.~V.}\ \bibnamefont
  {Mazurenko}},\ }\href {https://doi.org/10.1103/PhysRevB.98.174411} {\bibfield
   {journal} {\bibinfo  {journal} {Physical Review B}\ }\textbf {\bibinfo
  {volume} {98}},\ \bibinfo {pages} {174411} (\bibinfo {year}
  {2018})}\BibitemShut {NoStop}%
\bibitem [{\citenamefont {Singh}\ and\ \citenamefont {Han}(2019)}]{Singh.2019}%
  \BibitemOpen
  \bibfield  {author} {\bibinfo {author} {\bibfnamefont {V.~K.}\ \bibnamefont
  {Singh}}\ and\ \bibinfo {author} {\bibfnamefont {J.~H.}\ \bibnamefont
  {Han}},\ }\href {https://doi.org/10.1103/PhysRevB.99.174426} {\bibfield
  {journal} {\bibinfo  {journal} {Physical Review B}\ }\textbf {\bibinfo
  {volume} {99}},\ \bibinfo {pages} {174426} (\bibinfo {year}
  {2019})}\BibitemShut {NoStop}%
\bibitem [{\citenamefont {Kwon}\ \emph {et~al.}(2019)\citenamefont {Kwon},
  \citenamefont {Kim}, \citenamefont {Lee},\ and\ \citenamefont
  {Won}}]{Kwon.2019}%
  \BibitemOpen
  \bibfield  {author} {\bibinfo {author} {\bibfnamefont {H.~Y.}\ \bibnamefont
  {Kwon}}, \bibinfo {author} {\bibfnamefont {N.~J.}\ \bibnamefont {Kim}},
  \bibinfo {author} {\bibfnamefont {C.~K.}\ \bibnamefont {Lee}},\ and\ \bibinfo
  {author} {\bibfnamefont {C.}~\bibnamefont {Won}},\ }\href
  {https://doi.org/10.1103/PhysRevB.99.024423} {\bibfield  {journal} {\bibinfo
  {journal} {Physical Review B}\ }\textbf {\bibinfo {volume} {99}},\ \bibinfo
  {pages} {024423} (\bibinfo {year} {2019})}\BibitemShut {NoStop}%
\bibitem [{\citenamefont {Salcedo-Gallo}\ \emph {et~al.}(2020)\citenamefont
  {Salcedo-Gallo}, \citenamefont {Galindo-González},\ and\ \citenamefont
  {Restrepo-Parra}}]{SalcedoGallo.2020}%
  \BibitemOpen
  \bibfield  {author} {\bibinfo {author} {\bibfnamefont {J.}~\bibnamefont
  {Salcedo-Gallo}}, \bibinfo {author} {\bibfnamefont {C.}~\bibnamefont
  {Galindo-González}},\ and\ \bibinfo {author} {\bibfnamefont
  {E.}~\bibnamefont {Restrepo-Parra}},\ }\href
  {https://doi.org/https://doi.org/10.1016/j.jmmm.2020.166482} {\bibfield
  {journal} {\bibinfo  {journal} {Journal of Magnetism and Magnetic Materials}\
  }\textbf {\bibinfo {volume} {501}},\ \bibinfo {pages} {166482} (\bibinfo
  {year} {2020})}\BibitemShut {NoStop}%
\bibitem [{\citenamefont {Wang}\ \emph {et~al.}(2020)\citenamefont {Wang},
  \citenamefont {Wei}, \citenamefont {Yuan}, \citenamefont {Tian},
  \citenamefont {Cao}, \citenamefont {Zhao}, \citenamefont {Zhang},
  \citenamefont {Zhou}, \citenamefont {Song}, \citenamefont {Xue},\ and\
  \citenamefont {Yang}}]{Wang.2020}%
  \BibitemOpen
  \bibfield  {author} {\bibinfo {author} {\bibfnamefont {D.}~\bibnamefont
  {Wang}}, \bibinfo {author} {\bibfnamefont {S.}~\bibnamefont {Wei}}, \bibinfo
  {author} {\bibfnamefont {A.}~\bibnamefont {Yuan}}, \bibinfo {author}
  {\bibfnamefont {F.}~\bibnamefont {Tian}}, \bibinfo {author} {\bibfnamefont
  {K.}~\bibnamefont {Cao}}, \bibinfo {author} {\bibfnamefont {Q.}~\bibnamefont
  {Zhao}}, \bibinfo {author} {\bibfnamefont {Y.}~\bibnamefont {Zhang}},
  \bibinfo {author} {\bibfnamefont {C.}~\bibnamefont {Zhou}}, \bibinfo {author}
  {\bibfnamefont {X.}~\bibnamefont {Song}}, \bibinfo {author} {\bibfnamefont
  {D.}~\bibnamefont {Xue}},\ and\ \bibinfo {author} {\bibfnamefont
  {S.}~\bibnamefont {Yang}},\ }\href
  {https://doi.org/https://doi.org/10.1002/advs.202000566} {\bibfield
  {journal} {\bibinfo  {journal} {Advanced Science}\ }\textbf {\bibinfo
  {volume} {7}},\ \bibinfo {pages} {2000566} (\bibinfo {year}
  {2020})}\BibitemShut {NoStop}%
\bibitem [{\citenamefont {Kawaguchi}\ \emph {et~al.}(2021)\citenamefont
  {Kawaguchi}, \citenamefont {Tanabe}, \citenamefont {Yamada}, \citenamefont
  {Sawa}, \citenamefont {Hasegawa}, \citenamefont {Hayashi},\ and\
  \citenamefont {Nakatani}}]{Kawaguchi.2021}%
  \BibitemOpen
  \bibfield  {author} {\bibinfo {author} {\bibfnamefont {M.}~\bibnamefont
  {Kawaguchi}}, \bibinfo {author} {\bibfnamefont {K.}~\bibnamefont {Tanabe}},
  \bibinfo {author} {\bibfnamefont {K.}~\bibnamefont {Yamada}}, \bibinfo
  {author} {\bibfnamefont {T.}~\bibnamefont {Sawa}}, \bibinfo {author}
  {\bibfnamefont {S.}~\bibnamefont {Hasegawa}}, \bibinfo {author}
  {\bibfnamefont {M.}~\bibnamefont {Hayashi}},\ and\ \bibinfo {author}
  {\bibfnamefont {Y.}~\bibnamefont {Nakatani}},\ }\href
  {https://doi.org/10.1038/s41524-020-00485-2} {\bibfield  {journal} {\bibinfo
  {journal} {npj Computational Materials}\ }\textbf {\bibinfo {volume} {7}},\
  \bibinfo {pages} {20} (\bibinfo {year} {2021})}\BibitemShut {NoStop}%
\bibitem [{\citenamefont {Mamada}\ \emph {et~al.}(2021)\citenamefont {Mamada},
  \citenamefont {Mizumaki}, \citenamefont {Akai},\ and\ \citenamefont
  {Aonishi}}]{Mamada.2021}%
  \BibitemOpen
  \bibfield  {author} {\bibinfo {author} {\bibfnamefont {N.}~\bibnamefont
  {Mamada}}, \bibinfo {author} {\bibfnamefont {M.}~\bibnamefont {Mizumaki}},
  \bibinfo {author} {\bibfnamefont {I.}~\bibnamefont {Akai}},\ and\ \bibinfo
  {author} {\bibfnamefont {T.}~\bibnamefont {Aonishi}},\ }\href
  {https://doi.org/10.7566/jpsj.90.014705} {\bibfield  {journal} {\bibinfo
  {journal} {Journal of the Physical Society of Japan}\ }\textbf {\bibinfo
  {volume} {90}},\ \bibinfo {pages} {014705} (\bibinfo {year}
  {2021})}\BibitemShut {NoStop}%
\bibitem [{\citenamefont {Talapatra}\ \emph {et~al.}(2022)\citenamefont
  {Talapatra}, \citenamefont {Gajera}, \citenamefont {P}, \citenamefont
  {Chelvane},\ and\ \citenamefont {Mohanty}}]{Talapatra.2022}%
  \BibitemOpen
  \bibfield  {author} {\bibinfo {author} {\bibfnamefont {A.}~\bibnamefont
  {Talapatra}}, \bibinfo {author} {\bibfnamefont {U.}~\bibnamefont {Gajera}},
  \bibinfo {author} {\bibfnamefont {S.~P.}\ \bibnamefont {P}}, \bibinfo
  {author} {\bibfnamefont {J.~A.}\ \bibnamefont {Chelvane}},\ and\ \bibinfo
  {author} {\bibfnamefont {J.~R.}\ \bibnamefont {Mohanty}},\ }\href
  {https://doi.org/10.1021/acsami.2c12848} {\bibfield  {journal} {\bibinfo
  {journal} {ACS Applied Materials \& Interfaces}\ }\textbf {\bibinfo {volume}
  {14}},\ \bibinfo {pages} {50318} (\bibinfo {year} {2022})}\BibitemShut
  {NoStop}%
\bibitem [{\citenamefont {Kwon}\ \emph {et~al.}(2020)\citenamefont {Kwon},
  \citenamefont {Yoon}, \citenamefont {Lee}, \citenamefont {Chen},
  \citenamefont {Liu}, \citenamefont {Schmid}, \citenamefont {Wu},
  \citenamefont {Choi},\ and\ \citenamefont {Won}}]{Kwon.2020}%
  \BibitemOpen
  \bibfield  {author} {\bibinfo {author} {\bibfnamefont {H.~Y.}\ \bibnamefont
  {Kwon}}, \bibinfo {author} {\bibfnamefont {H.~G.}\ \bibnamefont {Yoon}},
  \bibinfo {author} {\bibfnamefont {C.}~\bibnamefont {Lee}}, \bibinfo {author}
  {\bibfnamefont {G.}~\bibnamefont {Chen}}, \bibinfo {author} {\bibfnamefont
  {K.}~\bibnamefont {Liu}}, \bibinfo {author} {\bibfnamefont {A.~K.}\
  \bibnamefont {Schmid}}, \bibinfo {author} {\bibfnamefont {Y.~Z.}\
  \bibnamefont {Wu}}, \bibinfo {author} {\bibfnamefont {J.~W.}\ \bibnamefont
  {Choi}},\ and\ \bibinfo {author} {\bibfnamefont {C.}~\bibnamefont {Won}},\
  }\href {https://doi.org/10.1126/sciadv.abb0872} {\bibfield  {journal}
  {\bibinfo  {journal} {Science Advances}\ }\textbf {\bibinfo {volume} {6}},\
  \bibinfo {pages} {eabb0872} (\bibinfo {year} {2020})}\BibitemShut {NoStop}%
\bibitem [{\citenamefont {LeCun}\ \emph {et~al.}(1989)\citenamefont {LeCun},
  \citenamefont {Boser}, \citenamefont {Denker}, \citenamefont {Henderson},
  \citenamefont {Howard}, \citenamefont {Hubbard},\ and\ \citenamefont
  {Jackel}}]{LeCun.1989}%
  \BibitemOpen
  \bibfield  {author} {\bibinfo {author} {\bibfnamefont {Y.}~\bibnamefont
  {LeCun}}, \bibinfo {author} {\bibfnamefont {B.}~\bibnamefont {Boser}},
  \bibinfo {author} {\bibfnamefont {J.~S.}\ \bibnamefont {Denker}}, \bibinfo
  {author} {\bibfnamefont {D.}~\bibnamefont {Henderson}}, \bibinfo {author}
  {\bibfnamefont {R.~E.}\ \bibnamefont {Howard}}, \bibinfo {author}
  {\bibfnamefont {W.}~\bibnamefont {Hubbard}},\ and\ \bibinfo {author}
  {\bibfnamefont {L.~D.}\ \bibnamefont {Jackel}},\ }\href
  {https://doi.org/10.1162/neco.1989.1.4.541} {\bibfield  {journal} {\bibinfo
  {journal} {Neural Computation}\ }\textbf {\bibinfo {volume} {1}},\ \bibinfo
  {pages} {541} (\bibinfo {year} {1989})}\BibitemShut {NoStop}%
\bibitem [{\citenamefont {LeCun}\ \emph {et~al.}(2015)\citenamefont {LeCun},
  \citenamefont {Bengio},\ and\ \citenamefont {Hinton}}]{LeCun.2015}%
  \BibitemOpen
  \bibfield  {author} {\bibinfo {author} {\bibfnamefont {Y.}~\bibnamefont
  {LeCun}}, \bibinfo {author} {\bibfnamefont {Y.}~\bibnamefont {Bengio}},\ and\
  \bibinfo {author} {\bibfnamefont {G.}~\bibnamefont {Hinton}},\ }\href
  {https://doi.org/10.1038/nature14539} {\bibfield  {journal} {\bibinfo
  {journal} {Nature}\ }\textbf {\bibinfo {volume} {521}},\ \bibinfo {pages}
  {436} (\bibinfo {year} {2015})}\BibitemShut {NoStop}%
\bibitem [{\citenamefont {Krizhevsky}\ \emph {et~al.}(2017)\citenamefont
  {Krizhevsky}, \citenamefont {Sutskever},\ and\ \citenamefont
  {Hinton}}]{Krizhevsky.2017}%
  \BibitemOpen
  \bibfield  {author} {\bibinfo {author} {\bibfnamefont {A.}~\bibnamefont
  {Krizhevsky}}, \bibinfo {author} {\bibfnamefont {I.}~\bibnamefont
  {Sutskever}},\ and\ \bibinfo {author} {\bibfnamefont {G.~E.}\ \bibnamefont
  {Hinton}},\ }\href {https://doi.org/10.1145/3065386} {\bibfield  {journal}
  {\bibinfo  {journal} {Communications of the ACM}\ }\textbf {\bibinfo {volume}
  {60}},\ \bibinfo {pages} {84} (\bibinfo {year} {2017})}\BibitemShut {NoStop}%
\bibitem [{\citenamefont {Shorten}\ and\ \citenamefont
  {Khoshgoftaar}(2019)}]{Shorten.2019}%
  \BibitemOpen
  \bibfield  {author} {\bibinfo {author} {\bibfnamefont {C.}~\bibnamefont
  {Shorten}}\ and\ \bibinfo {author} {\bibfnamefont {T.~M.}\ \bibnamefont
  {Khoshgoftaar}},\ }\bibfield  {journal} {\bibinfo  {journal} {Journal of Big
  Data}\ }\textbf {\bibinfo {volume} {6}},\ \href
  {https://doi.org/10.1186/s40537-019-0197-0} {10.1186/s40537-019-0197-0}
  (\bibinfo {year} {2019})\BibitemShut {NoStop}%
\bibitem [{\citenamefont {Hinton}\ \emph {et~al.}(2012)\citenamefont {Hinton},
  \citenamefont {Srivastava}, \citenamefont {Krizhevsky}, \citenamefont
  {Sutskever},\ and\ \citenamefont {Salakhutdinov}}]{Hinton.2012}%
  \BibitemOpen
  \bibfield  {author} {\bibinfo {author} {\bibfnamefont {G.~E.}\ \bibnamefont
  {Hinton}}, \bibinfo {author} {\bibfnamefont {N.}~\bibnamefont {Srivastava}},
  \bibinfo {author} {\bibfnamefont {A.}~\bibnamefont {Krizhevsky}}, \bibinfo
  {author} {\bibfnamefont {I.}~\bibnamefont {Sutskever}},\ and\ \bibinfo
  {author} {\bibfnamefont {R.~R.}\ \bibnamefont {Salakhutdinov}},\ }\href@noop
  {} {\bibfield  {journal} {\bibinfo  {journal} {arXiv:1207.0580 [cs]}\ }
  (\bibinfo {year} {2012})},\ \bibinfo {note} {arXiv: 1207.0580}\BibitemShut
  {NoStop}%
\bibitem [{\citenamefont {Chollet}(2017)}]{Chollet.2017}%
  \BibitemOpen
  \bibfield  {author} {\bibinfo {author} {\bibfnamefont {F.}~\bibnamefont
  {Chollet}},\ }\href@noop {} {\emph {\bibinfo {title} {Deep {Learning} with
  {Python}}}},\ \bibinfo {edition} {1st}\ ed.\ (\bibinfo  {publisher} {Manning
  Publications},\ \bibinfo {year} {2017})\BibitemShut {NoStop}%
\bibitem [{\citenamefont {Vértesy}\ and\ \citenamefont
  {Tomáš}(2003)}]{Vertesy.2003}%
  \BibitemOpen
  \bibfield  {author} {\bibinfo {author} {\bibfnamefont {G.}~\bibnamefont
  {Vértesy}}\ and\ \bibinfo {author} {\bibfnamefont {I.}~\bibnamefont
  {Tomáš}},\ }\href {https://doi.org/10.1063/1.1559434} {\bibfield  {journal}
  {\bibinfo  {journal} {Journal of Applied Physics}\ }\textbf {\bibinfo
  {volume} {93}},\ \bibinfo {pages} {4040} (\bibinfo {year}
  {2003})}\BibitemShut {NoStop}%
\bibitem [{\citenamefont {Lemesh}\ \emph {et~al.}(2017)\citenamefont {Lemesh},
  \citenamefont {B\"uttner},\ and\ \citenamefont {Beach}}]{Lemesh.2017}%
  \BibitemOpen
  \bibfield  {author} {\bibinfo {author} {\bibfnamefont {I.}~\bibnamefont
  {Lemesh}}, \bibinfo {author} {\bibfnamefont {F.}~\bibnamefont {B\"uttner}},\
  and\ \bibinfo {author} {\bibfnamefont {G.~S.~D.}\ \bibnamefont {Beach}},\
  }\href {https://doi.org/10.1103/PhysRevB.95.174423} {\bibfield  {journal}
  {\bibinfo  {journal} {Physical Review B}\ }\textbf {\bibinfo {volume} {95}},\
  \bibinfo {pages} {174423} (\bibinfo {year} {2017})}\BibitemShut {NoStop}%
\bibitem [{\citenamefont {Chen}\ \emph
  {et~al.}(2022{\natexlab{b}})\citenamefont {Chen}, \citenamefont {Chue},
  \citenamefont {Kong}, \citenamefont {Tan}, \citenamefont {Tan},\ and\
  \citenamefont {Soumyanarayanan}}]{Chen.2022a}%
  \BibitemOpen
  \bibfield  {author} {\bibinfo {author} {\bibfnamefont {X.}~\bibnamefont
  {Chen}}, \bibinfo {author} {\bibfnamefont {E.}~\bibnamefont {Chue}}, \bibinfo
  {author} {\bibfnamefont {J.~F.}\ \bibnamefont {Kong}}, \bibinfo {author}
  {\bibfnamefont {H.~R.}\ \bibnamefont {Tan}}, \bibinfo {author} {\bibfnamefont
  {H.~K.}\ \bibnamefont {Tan}},\ and\ \bibinfo {author} {\bibfnamefont
  {A.}~\bibnamefont {Soumyanarayanan}},\ }\href
  {https://doi.org/10.1103/physrevapplied.17.044039} {\bibfield  {journal}
  {\bibinfo  {journal} {Physical Review Applied}\ }\textbf {\bibinfo {volume}
  {17}},\ \bibinfo {pages} {044039} (\bibinfo {year}
  {2022}{\natexlab{b}})}\BibitemShut {NoStop}%
\bibitem [{\citenamefont {Legrand}\ \emph {et~al.}(2017)\citenamefont
  {Legrand}, \citenamefont {Maccariello}, \citenamefont {Reyren}, \citenamefont
  {Garcia}, \citenamefont {Moutafis}, \citenamefont {Moreau-Luchaire},
  \citenamefont {Collin}, \citenamefont {Bouzehouane}, \citenamefont {Cros},\
  and\ \citenamefont {Fert}}]{Legrand.2017}%
  \BibitemOpen
  \bibfield  {author} {\bibinfo {author} {\bibfnamefont {W.}~\bibnamefont
  {Legrand}}, \bibinfo {author} {\bibfnamefont {D.}~\bibnamefont
  {Maccariello}}, \bibinfo {author} {\bibfnamefont {N.}~\bibnamefont {Reyren}},
  \bibinfo {author} {\bibfnamefont {K.}~\bibnamefont {Garcia}}, \bibinfo
  {author} {\bibfnamefont {C.}~\bibnamefont {Moutafis}}, \bibinfo {author}
  {\bibfnamefont {C.}~\bibnamefont {Moreau-Luchaire}}, \bibinfo {author}
  {\bibfnamefont {S.}~\bibnamefont {Collin}}, \bibinfo {author} {\bibfnamefont
  {K.}~\bibnamefont {Bouzehouane}}, \bibinfo {author} {\bibfnamefont
  {V.}~\bibnamefont {Cros}},\ and\ \bibinfo {author} {\bibfnamefont
  {A.}~\bibnamefont {Fert}},\ }\href
  {https://doi.org/10.1021/acs.nanolett.7b00649} {\bibfield  {journal}
  {\bibinfo  {journal} {Nano Letters}\ }\textbf {\bibinfo {volume} {17}},\
  \bibinfo {pages} {2703} (\bibinfo {year} {2017})}\BibitemShut {NoStop}%
\bibitem [{\citenamefont {Juge}\ \emph {et~al.}(2019)\citenamefont {Juge},
  \citenamefont {Je}, \citenamefont {Chaves}, \citenamefont {Buda-Prejbeanu},
  \citenamefont {Pe{\~{n}}a-Garcia}, \citenamefont {Nath}, \citenamefont
  {Miron}, \citenamefont {Rana}, \citenamefont {Aballe}, \citenamefont
  {Foerster}, \citenamefont {Genuzio}, \citenamefont {Menteş}, \citenamefont
  {Locatelli}, \citenamefont {Maccherozzi}, \citenamefont {Dhesi},
  \citenamefont {Belmeguenai}, \citenamefont {Roussign{\'{e}}}, \citenamefont
  {Auffret}, \citenamefont {Pizzini}, \citenamefont {Gaudin}, \citenamefont
  {Vogel},\ and\ \citenamefont {Boulle}}]{Juge.2019}%
  \BibitemOpen
  \bibfield  {author} {\bibinfo {author} {\bibfnamefont {R.}~\bibnamefont
  {Juge}}, \bibinfo {author} {\bibfnamefont {S.-G.}\ \bibnamefont {Je}},
  \bibinfo {author} {\bibfnamefont {D.~D.~S.}\ \bibnamefont {Chaves}}, \bibinfo
  {author} {\bibfnamefont {L.~D.}\ \bibnamefont {Buda-Prejbeanu}}, \bibinfo
  {author} {\bibfnamefont {J.}~\bibnamefont {Pe{\~{n}}a-Garcia}}, \bibinfo
  {author} {\bibfnamefont {J.}~\bibnamefont {Nath}}, \bibinfo {author}
  {\bibfnamefont {I.~M.}\ \bibnamefont {Miron}}, \bibinfo {author}
  {\bibfnamefont {K.~G.}\ \bibnamefont {Rana}}, \bibinfo {author}
  {\bibfnamefont {L.}~\bibnamefont {Aballe}}, \bibinfo {author} {\bibfnamefont
  {M.}~\bibnamefont {Foerster}}, \bibinfo {author} {\bibfnamefont
  {F.}~\bibnamefont {Genuzio}}, \bibinfo {author} {\bibfnamefont {T.~O.}\
  \bibnamefont {Menteş}}, \bibinfo {author} {\bibfnamefont {A.}~\bibnamefont
  {Locatelli}}, \bibinfo {author} {\bibfnamefont {F.}~\bibnamefont
  {Maccherozzi}}, \bibinfo {author} {\bibfnamefont {S.~S.}\ \bibnamefont
  {Dhesi}}, \bibinfo {author} {\bibfnamefont {M.}~\bibnamefont {Belmeguenai}},
  \bibinfo {author} {\bibfnamefont {Y.}~\bibnamefont {Roussign{\'{e}}}},
  \bibinfo {author} {\bibfnamefont {S.}~\bibnamefont {Auffret}}, \bibinfo
  {author} {\bibfnamefont {S.}~\bibnamefont {Pizzini}}, \bibinfo {author}
  {\bibfnamefont {G.}~\bibnamefont {Gaudin}}, \bibinfo {author} {\bibfnamefont
  {J.}~\bibnamefont {Vogel}},\ and\ \bibinfo {author} {\bibfnamefont
  {O.}~\bibnamefont {Boulle}},\ }\href
  {https://doi.org/10.1103/PhysRevApplied.12.044007} {\bibfield  {journal}
  {\bibinfo  {journal} {Physical Review Applied}\ }\textbf {\bibinfo {volume}
  {12}},\ \bibinfo {pages} {044007} (\bibinfo {year} {2019})}\BibitemShut
  {NoStop}%
\bibitem [{\citenamefont {Chen}\ \emph {et~al.}(2023)\citenamefont {Chen},
  \citenamefont {Tai}, \citenamefont {Tan}, \citenamefont {Tan}, \citenamefont
  {Lim}, \citenamefont {Ho},\ and\ \citenamefont
  {Soumyanarayanan}}]{Chen.2023}%
  \BibitemOpen
  \bibfield  {author} {\bibinfo {author} {\bibfnamefont {X.}~\bibnamefont
  {Chen}}, \bibinfo {author} {\bibfnamefont {T.}~\bibnamefont {Tai}}, \bibinfo
  {author} {\bibfnamefont {H.}~\bibnamefont {Tan}}, \bibinfo {author}
  {\bibfnamefont {H.}~\bibnamefont {Tan}}, \bibinfo {author} {\bibfnamefont
  {R.}~\bibnamefont {Lim}}, \bibinfo {author} {\bibfnamefont {P.}~\bibnamefont
  {Ho}},\ and\ \bibinfo {author} {\bibfnamefont {A.}~\bibnamefont
  {Soumyanarayanan}},\ }\href@noop {} {\bibfield  {journal} {\bibinfo
  {journal} {ArXiv e-prints}\ } (\bibinfo {year} {2023})},\ \Eprint
  {https://arxiv.org/abs/2301.07327} {arXiv:2301.07327 [cond-mat.mtrl-sci]}
  \BibitemShut {NoStop}%
\bibitem [{\citenamefont {Otsu}(1979)}]{Otsu.1979}%
  \BibitemOpen
  \bibfield  {author} {\bibinfo {author} {\bibfnamefont {N.}~\bibnamefont
  {Otsu}},\ }\href {https://doi.org/10.1109/TSMC.1979.4310076} {\bibfield
  {journal} {\bibinfo  {journal} {IEEE Transactions on Systems, Man, and
  Cybernetics}\ }\textbf {\bibinfo {volume} {9}},\ \bibinfo {pages} {62}
  (\bibinfo {year} {1979})}\BibitemShut {NoStop}%
\end{thebibliography}%

\noindent \begin{center}
{\small{}\rule[0.5ex]{0.6\columnwidth}{0.5pt}}{\small\par}
\par\end{center}

\end{document}